\documentclass{tlp}
 
\usepackage{eepic}
\usepackage{amsmath}
\usepackage{oz}
\usepackage{reflp}

\fuzzcompatible

\newcommand\reflptool{\mbox{\tt Marvin}}

\title{A Refinement Calculus for Logic Programs}
\author[I. Hayes et al.]
{IAN HAYES, ROBERT COLVIN, DAVID HEMER and PAUL STROOPER\\
 School of Information Technology and Electrical Engineering,\\
The University of Queensland, Australia
\and   RAY NICKSON\\
School of Mathematical and Computing Sciences,\\
Victoria University of Wellington, New Zealand}

\begin{document}

\maketitle

\begin{abstract}
Existing refinement calculi provide frameworks for the stepwise development of
imperative programs from specifications.
This paper presents a refinement calculus for deriving logic programs.
The calculus contains a wide-spectrum logic programming language, including
executable constructs such as sequential conjunction, disjunction, and
existential quantification, as well as specification constructs
such as general predicates, assumptions and universal quantification.
A declarative semantics is defined for this wide-spectrum language based on
{\em executions}. Executions are partial functions from states to states,
where a state is represented as a set of bindings.
The semantics is used to define the meaning of programs and
specifications, including parameters and recursion.
To complete the calculus, a notion of correctness-preserving refinement over
programs in the wide-spectrum language is defined
and refinement laws for developing programs are introduced.
The refinement calculus is illustrated using example derivations and
prototype tool support is discussed.
\end{abstract}

\section{Introduction} 
Our goal is to provide a method for the
systematic development of logic programs from specifications.  We
follow a refinement calculus approach \cite{CPPR,SSR,ATBfSRatPC,PfS},
which provides a framework for the stepwise development of imperative
programs from specifications.  It makes use of a wide-spectrum
language that includes both specification and programming language
constructs.  This allows a specification to be refined, step by step,
to program code within a single language.  The refinement steps are
performed using refinement laws that have been proven correct within
the framework. Once they are proven correct, they can be safely
applied in any refinement, although in some cases their application
involves proof obligations that must be discharged.  The {\em
  programs} produced during the intermediate steps of the refinement
process may contain specification constructs as components, and hence
may not be {\em code} suitable for execution.
The refinement is completed when the program contains only executable 
constructs.

We define a refinement calculus for logic programming.
The calculus contains a wide-spectrum logic
programming language, including executable conjunction, disjunction,
and existential quantification, as well as specification constructs
such as general predicates, assumptions and universal quantification,
which are not in general executable.
General predicates allow the effect of a program to be specified via
properties expressed as logical formulae.
Assumptions represent information about the context in which
a program fragment will execute.
An implementation is obliged to produce the specified result only if
its assumptions are satisfied by the context.
The language also supports parametrised procedures and recursion.

We define a declarative semantics for the wide-spectrum language in terms of
{\em executions}, which are partial functions from initial to final states.
A state in turn is represented as a set of bindings, where each binding
is a mapping from variables to values.
As is traditionally the case with logic programs, we consider only
executions where the set of bindings in the final state are a subset of the
bindings in the initial state.
To complete the calculus, we define a notion of correctness-preserving
refinement over programs in the wide-spectrum language
and a set of refinement laws.
The declarative nature of the semantics means that we do not
distinguish a program that produces an answer once from a program that
produces the same answer multiple times, nor do we consider the order in which
answers are produced.
Our semantics makes use of a least fixed point semantics that equates
non-termination with the least defined program, $\abort$.

Section~2 of this paper presents related work.
Section~3 summarizes the wide-spectrum logic programming language.
Section~4 gives the basic definitions necessary for our formal semantics.
Section~5 presents the semantics of the base language in terms of executions.
Section~6 defines our notion of refinement.
Section~7 gives the machinery for dealing with procedures and parameters.
Section~8 discusses the semantics of recursion.
Section~9 discusses  refinement laws 
and presents a small example.
Section~10 presents an extended example, which is the refinement of a
program for the N-queens problem.
Section~11 discusses \reflptool{}, a prototype tool that supports the
refinement calculus.

Our semantics is described using the mathematical notation of the
Z specification
language~\cite{SPIV:Z-REF}, except that we write sequences within 
square brackets to be consistent with logic programming notation.
No familiarity with $Z$ is assumed.
A number of properties of the semantics are presented here without proof;
the proofs of these properties are presented in \cite{tr00-30}.

\section{Related work}

Traditionally, the refinement calculus has been used to develop imperative
programs from specifications \cite{CPPR,SSR,ATBfSRatPC,PfS}.  The increase in
expressive power of logic programming languages, when compared with imperative
languages, leads to a reduced conceptual gap between a problem and its
solution, which means that fewer development steps are required during
refinement.  An additional advantage of logic programming languages over
procedural languages is their  simpler, cleaner semantics, which leads to
simpler proofs of the refinement steps.  Finally, the higher expressive level
of logic programming languages means that the individual refinement steps
typically achieve more.

There have been previous proposals for developing a refinement
calculus for declarative languages.  A refinement calculus for
functional programming languages is discussed by Ward \shortcite{Ward94}.
Kok \shortcite{Kok90} has applied the refinement calculus to logic
programming languages, but his approach is quite different to ours.
Rather than defining a wide-spectrum logic programming language, Kok
embeds logic programs into a more traditional refinement calculus
framework and notation.

There have been several proposals for the constructive development of
logic programs, for example in Jacquet \shortcite{Jacquet93}.
Much of this work has focused on
program transformations or equivalence transformations from a
first-order logic specification \cite{Clark78,Hogger81}.  Read and
Kazmierczak \shortcite{Read91} propose a stepwise development of
modular logic programs from first-order specifications, based on three
refinement steps that are much coarser than the refinement steps
proposed in this paper.  This leaves most of the work to be done in
discharging the proof obligations for the refinement steps, for which
they provide little guidance.
Another approach to constructing logic programs is through $schemata$
\cite{Marakakis:97}.  A logic program is designed through the application
of common algorithmic structures.  The designer chooses which program structure
is most suitable to a task based on the data types in question.
As such, the focus of this method is to aid the design of large programs.
The refinement steps and corresponding verification proofs
are therefore much larger.
Deductive logic program synthesis
\cite{LauDeville:94} is probably the most similar to the refinement calculus
approach.
In deductive synthesis,
a specification is successively transformed using
synthesis laws proven in an underlying framework 
(typically first-order logic).

Two key aspects that make the refinement calculus approach different
from these other proposals are the smaller refinement steps and the
use of assumptions. Refinement steps are performed by applying
individual refinement laws that have been proven correct within the
framework; these laws have a much smaller granularity than the
refinement steps in the other approaches. Applying these refinement
laws may introduce proof obligations that must be discharged, however
discharging these proof obligations is usually straightforward, and
can often be handled automatically with suitable tool support.  The
use of assumptions allows refinements to be proved correct with
respect to the context in which they appear.  To our knowledge, such
{\em refinements in context} are novel in the logic programming
community.

Deville \shortcite{Deville90} introduces a  systematic program development
method for Prolog that incorporates assumptions and types similar to ours.  The
main difference is that Deville's approach to program development is mostly
informal, whereas our approach is fully formal.  A second distinction is that
Deville's approach concentrates on the development of individual procedures.
By using a wide-spectrum language, our approach blurs the distinction between a
logic description and a logic program.  For example, general predicates may
appear anywhere within a program, and the refinement rules allow them to be
transformed within that context.  Similarly, programming language constructs
may be used and transformed at any point.

The motivation for the work by Hoare \shortcite{Hoare:00} is to come up with
unifying theories for programming, which is quite different from
the motivation for our work.
However, the logic programming constructs he considers
and the semantics he uses are both very similar to the ones we use.

\section{Wide-Spectrum Language}

\label{reflp:wide-sp-l}
This section presents the wide-spectrum logic programming language
\cite{HNS:REFLP}, which combines both logic programming language and
specification language constructs.
It allows constructs that may not
be executable, similar to Back's \shortcite{CPPR}
inclusion of specification constructs in Dijkstra's imperative
language.  This has the benefit of allowing gradual refinement without
the need for notational changes during the refinement process.

The constructs in the language are
specifications (also called general predicates),
assumptions,
propositional operators,
quantifiers,
and
procedure calls.
A summary of the language
is shown in Figure~\ref{wide-spec-lang}.
\begin{figure}
\begin{center}
\begin{tabular}{rcl}
    $\Spec{P}$ & - & specification\\
    \{A\} & - & assumption \\
    $(c_1 \lor c_2)$ & - & disjunction \\
    $(c_1 \land c_2)$ & - & parallel conjunction \\
    $(c_1,c_2)$ & - & sequential conjunction \\
    ($\exists V @ c)$ & - &  existential quantification \\
    $(\forall V @ c)$ & - &  universal quantification \\
    $pc(t)$ & - & procedure call
\end{tabular}
\end{center}
\caption{Summary of wide-spectrum language commands.}
\label{wide-spec-lang}
\end{figure}

{\bf Specifications:} A specification $\Spec{P}$, where $P$ is a
predicate, represents a set of instantiations of the free variables of
the program that satisfy $P$.  For example, the command
$\Spec{X = 5 \lor X = 6}$ represents the set of instantiations $\{ 5,6 \}$ for
$X$,
and
the command $\Spec{V = fact(U)}$ specifies the set of pairs of $V$ and $U$
such that $V$ equals the factorial of $U$.
The specification $\fail$ is defined by:
\begin{zed}
  \fail == \SpecCmd(false)
\end{zed}
(We use the notation `$==$' to indicate a definition.)
The program
$\fail$ always computes an empty answer set, like Prolog's {\tt fail}.
The null command, $\Skip$, is defined by
\begin{zed}
  \Skip == \SpecCmd(true)
\end{zed}

{\bf Assumptions:}  An assumption $\{ A \}$, where $A$ is a predicate,
expresses a requirement on the context for a program fragment.
For example, we may augment our specification of the factorial example
by an assumption that $U$ has been instantiated to be an element of
the natural numbers before the factorial is evaluated:
$\Assert{U \in \nat},\Spec{V = fact(U)}$.
If assumptions about
the context are formally expressed, implementations may take advantage
of the assumptions, but need not establish (or indeed check) them.
If these assumptions do not hold, the program fragment may abort.
Aborting includes program behaviour such as nontermination
and abnormal termination due to
exceptions like division by zero, as well
as termination with arbitrary results.
We define the (worst possible) program $\abort$ by
\begin{zed}
  \abort == \AssertCmd(\false)
\end{zed}
Note that $\abort$ is quite different from the program $\fail$,
which never aborts, but has an empty solution set.

{\bf Propositional Operators:}
There are two forms of conjunction: a sequential form
$(c_1,c_2)$ where command $c_1$ is evaluated before $c_2$;
and a parallel version $(c_1 \land c_2)$ where $c_1$ and $c_2$
are evaluated independently and the intersection of their respective
results is formed.
The disjunction of two programs $(c_1 \lor c_2)$ computes the
union of the results of the two programs.  
We overload the symbols `$\land$' and `$\lor$' as both operators
on predicates and operators on commands.
Because, for example, the meanings of
$\Spec{P \land Q}$ and $\Spec{P} \prand \Spec{Q}$ are identical,
this does not usually cause confusion.

The following three programs illustrate the behaviour of sequential and parallel
conjunction, and show the difference between abortion and failure.
\[
  P1 == \Assert{X \neq 0}, \Spec{Y=1/X} \\
  P2 == \Spec{X \neq 0}, \Spec{Y=1/X} \\
  P3 == \Spec{X \neq 0} \pand \Spec{Y=1/X} \\
\]
If each of the three programs is executed from a state where $X=0$,
$P1$ will abort, while $P2$ will fail, producing an empty answer set.
The behaviour of $P3$ is also equivalent to abort, because the predicate
$Y=1/0$ is not defined.

{\bf Quantifiers:}
The existential quantifier $(\exists V@c)$ generalises disjunction,
computing the union of the results of command
$c$ for all possible values of $V$.
For example, the following may be part of the development of a
factorial procedure:
\[
  \exists U1,V1 @ \Spec{U1 = U-1 \land V1 = fact(U1) \land V = V1*U}
\]
Similarly, the universal quantifier $(\forall V@c)$
computes the intersection of the results of command $c$ for all
possible values of $V$.

{\bf Parametrised Commands and Procedures:}
We separate the definition of parametrised commands from procedures.
A parametrised command has the form
\[
  \pcmd(V,c)
\]
where $V$ is a variable, and $c$ is a wide-spectrum command.
This notation is similar is the lambda calculus notation,
$(\lambda V @ c)$,
and allows anonymous procedures to be used.

A procedure definition has the form
\[
  id \defs pc
\]
where $id$ is an identifier representing the name of the procedure,
and $pc$ is a parametrised command as defined above.
For example,
\[
  id \defs \pcmd(V,c)
\]
defines a procedure called $id$ with a formal
parameter $V$ and body $c$.
A parametrised command or a procedure identifier that is defined as a
parametrised command may be applied to an actual parameter term.
For example,
a call on the procedure $id$ is of the form $id(t)$, where
$t$ is a term: the actual parameter.

Our semantics only deals with procedures with a single parameter.
However, no generality is lost because multiple parameters
may be encoded using compound terms.
For example, a procedure to calculate factorials may be specified by
\[
  factorial \defs \pcmd({(U,V)},\Assert{U \in \nat},\Spec{V = fact(U)}) 
\]
which is an abbreviation for the following definition with a single
parameter $W$:
\[
  factorial \defs \pcmd(W,{(\exists U,V @ 
    \Spec{W = (U,V)},\Assert{U \in \nat},\Spec{V = fact(U)})})
\]


{\bf Commands:}
\label{def-cmd}
We define $Cmd$ to be the set of commands in our language, built
up from the constructs shown in Figure~\ref{wide-spec-lang}.


\section{Semantic Domains}
We begin our formal treatment of the semantics by defining the domains
over which our semantics of programs are given.

\subsection{Variables, values, and functors}
\label{def-var}
\label{def-val}
\label{def-fun}
We have fixed domains of {\em variables\/} ($Var$), {\em values} ($Val$),
and {\em functors} ($Fun$).
Elements of $Var$ represent the variables that can occur in programs.
This includes variables for which a program's answers give values,
variables that are bound by universal and existential quantifiers,
and variables used as formal parameters.
Values are the objects in the universe of discourse, denoted by ground
terms.
Functors represent the function symbols in our language that are used
to construct compound terms.
The arity of a functor is given by $\arity$, which is a total
function from functors
to natural numbers (notationally,
$\arity: Fun \fun \nat$).
Atoms are functors of arity zero.

If we restrict our interpretation of ground terms to the Herbrand
interpretation, as is typical in logic programming, $Val$ is
structured into atoms and compound terms.  But we allow more
structure than this; $Val$ may be structured according to any algebra,
taking into account whatever kind of term equality is appropriate for
the application under consideration.  For the purposes of this paper,
all we assume is that there is a function $\apply$,
which models the application of a function
to a sequence of values of length equal to the arity of the function,
resulting in a value.
We present the definition of $\apply$ as a $Z$ axiomatic definition;
the signature of $\apply$ is given above the line,
and the definition in the form of a predicate is given below the line.
In this case the signature states that $\apply$ is a (total)
function that takes a
functor, and returns a \emph{partial} function ($\pfun$) mapping sequences
of values to values.
\label{def-apply}
\begin{axdef}
  \apply: Fun \fun (\seq Val \pfun Val)
\where
  \forall f: Fun @ \dom(\apply~f) \subseteq \{ s: \seq Val | \# s = \arity~f \}
\end{axdef}
With this more general interpretation, given a functor $f$ of arity $n$,
`$\apply f$' may be undefined for some sequences of values of length $n$.
We write $def~E$ (or $def~P$) for the predicate that is true precisely
when the expression $E$ (or predicate $P$) is well-defined: that is, when
all function applications occurring within it are well-defined.  For
the Herbrand interpretation, or indeed for any other interpretation in
which all functions are total, $def~E = def~P = true$ for all
expressions $E$ and predicates $P$.

\subsection{Bindings, States and Predicates}
\label{def-bnd}
A {\em binding\/} is a total function, mapping every variable to a value.
\begin{zed}
  Bnd == Var \fun Val
\end{zed}
Each binding corresponds to a single ground answer to a Prolog-like query.
The mechanism for representing ``unbound'' variables is described below.

\label{def-state}
A {\em state\/} is a set of bindings.
\begin{zed}
  State == \power Bnd
\end{zed}
A state corresponds to our usual notion of a predicate with some
free variables, which is true or false once provided with a binding
for those variables, i.e., for a binding in the state.  Given a
predicate $P$, we write $\pred P$ to denote the set of bindings
satisfying $P$. 
For example:
\begin{eqnarray*}
  \pred{false} & = & \emptyset \\
  \pred{true}  & = & Bnd \\
  \pred{X=3}   & = & \{b:Bnd | b~X = 3\} \\
  \pred{X<Y}   & = & \{b:Bnd | b~X < b~Y\}
\end{eqnarray*}

A completely unbound variable is represented by a possibly infinite state
that has one binding to each element of $Val$.
For example, if we suppose that $Var$ contains just the variables
$\{X,Y,Z\}$, $Val$ is the set $\{0,1\}$, and $f$ has arity one and is total,
the set of solutions to the equation
$Y=f(X)$ is represented by the following set of bindings.
\[
\!\!\!\{ \{X \mapsto 0, Y \mapsto \apply~f~[ 0 ], Z \mapsto 0 \},  \\
        \{X \mapsto 0, Y \mapsto \apply~f~[ 0 ], Z \mapsto 1 \},  \\
        \{X \mapsto 1, Y \mapsto \apply~f~[ 1 ], Z \mapsto 0 \},  \\
        \{X \mapsto 1, Y \mapsto \apply~f~[ 1 ], Z \mapsto 1 \}  \}
\]
The expression $X \mapsto x$ constructs a pair of its operands;
it is equivalent to $(X,x)$.


We do not concern ourselves in this semantics with finite
or infinite representation of infinite states.  Our model allows a
successful execution that terminates in an infinite state.  Some
infinite states may correspond to Prolog
answer sets that can be finitely enumerated, and others may not.

\subsection{Terms}
\label{def-term}
A {\em Term\/} is a variable, or a functor with a (possibly empty)
finite sequence of terms (the arguments).
We model a term via the following data type constructor.
\begin{syntax}
  Term ::= varT \ldata Var \rdata | funT \ldata Fun \cross \seq Term \rdata
\end{syntax}
For any variable $V$, $varT(V)$ is a term,
and
if $f$ is a functor and $ts$ is a sequence of
terms, then $funT(f,ts)$ is a term.

A term may have a value when evaluated relative to some binding, or it
may be undefined if the term involves the incorrect application
of a function.
We define a partial function $\eval$ that
evaluates a term relative to a binding.
\label{def-eval}
\begin{axdef}
  \eval: Term \fun (Bnd \pfun Val)
\where
  \eval(varT(V)) = (\lambda b: Bnd @ b~V) \\
  \eval(funT(f,ts)) = \{ b: Bnd; vs: \seq Val | \\
\t2    (\forall t: \ran ts @ b \in \dom(\eval~t)) \land \\
\t2    vs = map (\lambda t: Term @ \eval~t~b)(ts) \land \\
\t2    vs \in \dom(\apply~f) \\
\t2    @ b \mapsto \apply~f~vs
      \}
\end{axdef}
The notation $\{x:T| P @ E\}$ describes
the set of values of the expression $E$, for each
$x$ of type $T$ for which $P$ holds
(when $P$ is $true$ we may omit it, i.e., $\{x:T @ E\}$).
Thus, $\{x:T | P @ x \mapsto E\}$ is the function (set of pairs)
whose domain is the set of $x:T$ satisfying predicate $P$, and which maps
each such $x$ to the corresponding value of expression $E$.
In the definition of $\eval$,
to evaluate a compound term $funT(f,ts)$
with respect to a binding $b$,
all of the terms in the sequence $ts$ must be able to be
evaluated with respect to $b$,
the resultant sequence of values $vs$ is defined by mapping
a function that evaluates a single term with respect to $b$ over the
sequence $ts$, and $vs$ must be in the
domain of $\apply~f$.  

For a term $t$, $\defined~t$ is the set of states for which
$t$ is defined for every binding in the state.
\label{def-defined}
\begin{axdef}
  \defined: Term \fun \power State
\where
  \defined~t =
    \{ s: State | s \subseteq \dom(\eval~t) \}
\end{axdef}

For a variable $V$, term $t$, and state $s$,
$\assign~V~t~s$ is the same as state $s$,
except that in each binding within $s$ the value of $V$
is replaced by the value of $t$ in that binding.
\label{def-assign}
\begin{axdef}
  \assign: Var \fun Term \fun State \pfun State
\where
  \assign~V~t = (\lambda s: \defined~t @
    \{ b: s @ b \oplus \{ V \mapsto \eval~t~b \} \})
\end{axdef}
In the definition, 
$f \oplus g$ stands for function $f$ overridden by function
$g$.
The expression $f \oplus g$ is the same as function $f$,
but with everything in the domain
of $g$ mapped to the same objects that $g$ maps them to 
(outside of the domain of $g$, $f \oplus g$ behaves like $f$).

For some term $t$, $\free t$ is the set of free variables in $t$:
\label{def-free}
\begin{axdef}
  \free: Term \fun \power Var
\where
  \free(varT(V)) = \{V\} \\
  \free(funT(f,ts)) = \stateuni \{ t: \ran ts @ \free t \}
\end{axdef}
Similarly, for a command $c$, $\free c$ defines
the set of free variables in $c$.
Since we do not formally define the predicates allowed in specifications
and assumptions here, we do not define $\free c$ formally.

\section{Program execution}
\label{SEC:execs}
\subsection{Executions}
\label{sec:execution:properties}
We define the semantics of our language in terms of {\em executions},
which are mappings from initial states to final states.
The mapping is partial because the program is only well-defined for
those initial states that guarantee satisfaction of all the program's
assumptions.
Executions satisfy three healthiness properties below. 
These properties restrict executions to model pure logic programs.

\begin{enumerate}

\item
\label{exec-per-binding-dom}
If a command is guaranteed to terminate from an initial state
$\pred{P}$ whose bindings all satisfy some predicate $P$, it must also
guarantee to terminate from all those initial states $\pred{P'}$,
where $P' \implies P$.  We thus require that any subset $s'$ of a
set $s$ that is in the domain of an execution $e$, is also in the
domain of $e$.
\[
        \forall s: \dom e @ (\forall s':\power s @ s' \in \dom e) \\
\]

In addition, if a command is guaranteed to terminate from initial
state $\pred P$ and it is also guaranteed to terminate from initial
state $\pred Q$, it must terminate from an initial state $\pred{P \lor Q}$.
Thus, if all sets in a set of states $ss$ are in the domain
of $e$, then their union is also in the domain of $e$.
\[
        \forall ss: \power (\dom e) @ \stateuni ss \in \dom e
\]
As we show in \cite{tr00-30}, these together are
equivalent to the fact that the domain of $e$ is the powerset of the
set of all bindings, $b$, such that the singleton set
$\{ b \}$ is in the domain of $e$.
\[
        \dom e = \power \{b:\ Bnd | \{ b \} \in \dom e\}
\]

\item
\label{exec-constraining}
Because of the constraining nature of logic programs
(command execution cannot decrease ``groundedness'')
for any state $s$ in the domain of an execution $e$, the set of
bindings in the output state $e(s)$ must be a subset of $s$.
\[
   \forall s: \dom e @ e(s) \subseteq s
\]
Consider a state with variables $X$ and $Y$ which can
take values in the set $\{ 0, 1 \}$ and the execution $e$
that represents the program $\Spec{X = Y}$. Intuitively the set of
initial states for a specification $\Spec{P}$ consists of those states
for which $P$ is defined. In this example the domain of $e$ consists
of all subsets of 
\begin{eqnarray*}
S & = & \{ \{ X \mapsto 0, Y \mapsto 0 \},
\{ X \mapsto 0, Y \mapsto 1 \}, \\
  &   & ~~\{ X \mapsto 1, Y \mapsto 0 \},
\{ X \mapsto 1, Y \mapsto 1 \} \}
\end{eqnarray*}
Given an initial state, the effect of executing $\Spec{X = Y}$ is to
filter out those bindings $b$ for which $X = Y$ is not true.  In
particular $e(S) = \{ \{ X \mapsto 0, Y \mapsto 0 \},\{ X \mapsto 1, Y
\mapsto 1 \} \} \subseteq S$.

\item
\label{exec-per-binding-range}
For a set of bindings $s$, the output set of bindings
consists of all those bindings $b$, such that the singleton set
$\{b\}$ is preserved by $e$.
\[
    \forall s: \dom e @ e(s) = \{b:s|e(\{b\}) = \{b\} \}
\]
For each of the singleton states in the example state $S$ above, 
\begin{eqnarray*}
&& e( \{ \{ X \mapsto 0, Y \mapsto 0 \}\}) = \{ \{ X \mapsto 0, Y \mapsto 0 \}\} \\
&& e( \{ \{ X \mapsto 0, Y \mapsto 1 \}\}) = \{ \} \\
&& e( \{ \{ X \mapsto 1, Y \mapsto 0 \}\}) = \{ \}\\
&& e( \{ \{ X \mapsto 1, Y \mapsto 1 \}\}) = \{ \{ X \mapsto 1, Y \mapsto 1 \}\}
\end{eqnarray*}
Thus $e(S)$ consists of the bindings which are individually
preserved by $e$.
\end{enumerate}

\noindent
We thus define:
\label{def-exec}
\begin{eqnarray}
  \setcounter{equation}{1}
  && Exec == \{ e: State \pfun State | \nonumber \\
  &&      \t2 \dom e = \power \{b:\ Bnd | \{ b \} \in \dom e\} \land \\
  &&      \t2 (\forall s: \dom e @ e(s) \subseteq s) \land \\
  &&      \t2 (\forall s: \dom e @ e(s) = \{b:s|e(\{b\}) = \{b\} \})
          \}
\end{eqnarray}

Note that Property~(\ref{exec-per-binding-dom}) implies that
$\emptyset \in \dom e$ for all executions $e$.  Also, from Property
(\ref{exec-constraining}), $e(\{b\})$ is either $\{b\}$ or $\emptyset$.
In \cite{tr00-30}, Property (3) is shown to be equivalent to either
of the following two properties.
\begin{eqnarray}
  && \forall s: \dom e @ e(s) = \stateuni \{b:s @ e(\{b\})\}
          \label{union-of-bindings} \\
  && \forall s: \dom e @ \forall s':\power s @ e(s') = e(s) \cap s'
          \label{intersection-with-subset} 
\end{eqnarray}
Property (\ref{union-of-bindings}) shows that an execution $e$
can be determined by considering the effect of the execution on
each singleton binding, and then forming the union of the results.
Property (\ref{intersection-with-subset}) shows that the result of
executing a command in a subset
$s'$ of some state $s$ is consistent with executing the
command in state $s$ and restricting the results to those in $s'$.
This property is similar to the property quoted by Hoare \shortcite{Hoare:00}
and attributed to He Jifeng, as one that characterises a pure logic program.
For example, Prolog's {\tt var} does not satisfy the property.

\subsection{Semantic function for commands}
We define the semantics of the commands in our language via a function
that takes a command and returns the corresponding execution.
\label{def-cexec}
\begin{axdef}
  \cexec: Cmd \fun Exec
\end{axdef}
The semantics of the basic commands (excluding procedure calls, which
are treated in Section~\ref{sec:procedure-calls})
is shown in Figure~\ref{fig-cexec-basic}.
In the remainder of this section, we explain the definitions.
In \cite{tr00-30}, we show that all executions constructed
using the definitions satisfy the healthiness properties of executions.

In Section~\ref{sec:procedures}, where we discuss procedures and
parameters, we extend the definition of $\cexec$ with an {\em
  environment}, which maps procedure identifiers to their
corresponding (parametrised) executions.  For simplicity, we first present the
semantics of the basic commands ignoring the environment.

\begin{figure}
\begin{center}
\begin{zed}
    \cexec(\SpecCmd(P)) = 
      (\lambda s:\power ( \predfn(def~P) ) @ s \cap \predfn(P)) \\
    \cexec(\AssertCmd(A)) =
      (\lambda s:\power ( \predfn(def~A \prand A) ) @ s) \\
    \cexec(\fail) = \cexec(\SpecCmd(false)) = (\lambda s:State @ \emptyset) \\
    \cexec(\abort) = \cexec(\AssertCmd(false)) = \{ \emptyset \mapsto \emptyset \} \\
    \cexec(\Skip) = \cexec(\SpecCmd(true)) = (\lambda s:State @ s) \\
    \cexec(c_1 \por c_2) = \cexec c_1 \lcup \cexec c_2 \\
    \cexec(c_1 \pand c_2) = \cexec c_1 \lcap \cexec c_2 \\
    \cexec(c_1 \sand c_2) = \cexec c_1 \comp \cexec c_2 \\
    \cexec(\pexists V \pat c) = \eex V (\cexec c) \\
    \cexec(\pforall V \pat c) = \eall V (\cexec c)
\end{zed}
\end{center}
\caption{Execution semantics of basic commands}
\label{fig-cexec-basic}
\end{figure}

\subsection{Specifications and assumptions}
A specification $\Spec{P}$ is defined for all states $s$ such that $P$ is defined
for all bindings in $s$; the result of executing specification $\Spec{P}$
consists of those bindings in $s$ that satisfy $P$.

An assumption $\{A\}$ is defined for all states $s$ such that $A$ is defined
and $A$ holds for all bindings in $s$; the result of executing assumption
$\{A\}$ has no effect (the set of bindings remains unchanged).

The definition for the special-case specification $\fail$ is the
constant function that returns the empty state, no matter what the
initial state is.
Hence for any command $c$, including $\abort$,
\[
  \fail \sand c = \fail
\]
because $\fail$ maps any state to the empty state.

The definition for the special-case assumption $\abort$ is the
function mapping the empty state to the empty state.
Hence for any command $c$,
\[
  \abort \sand c = \abort
\]
because the domain of $\abort$ contains only the empty state,
which it maps to the empty state.
Note that the empty state is preserved by any command, i.e.,
for any command $c$, $(\cexec~c)(\emptyset) = \emptyset$.

The definition of the special-case specification $\Skip$ is the
identity function on states.
Hence for any command $c$,
\[
  c \sand \Skip = c = \Skip \sand c .
\]

\subsection{Propositional operators}
Disjunction and parallel conjunction are defined as pointwise
union and intersection of the corresponding executions.
\label{def-lcap}
\begin{axdef}
  \_ \lcap \_: Exec \cross Exec \fun Exec \\
  \_ \lcup \_: Exec \cross Exec \fun Exec
\where
   (e_1 \lcap e_2) = 
      (\lambda s:\dom e_1 \cap \dom e_2 @
         (e_1~s) \cap (e_2~s)) \\
   (e_1 \lcup e_2) = 
      (\lambda s:\dom e_1 \cap \dom e_2 @
         (e_1~s) \cup (e_2~s))
\end{axdef}
For a conjunction ($c_1 \prand c_2)$, if a state $s$ is
mapped to $s'$ by $\cexec c_1$ and $s$ is mapped to $s''$ by $\cexec c_2$,
then $\cexec(c_1 \prand c_2)$ maps $s$ to $s' \cap s''$.  Disjunction
is similar, but gives the union of the resulting states instead of
intersection.
The command $\abort$ is a zero of both disjunction and parallel
conjunction,
$\Skip$ is the unit of parallel conjunction,
and
$\fail$ is the unit of disjunction.

Sequential conjunction $(c_1 \sand c_2)$ is defined
as function composition of the corresponding executions.
The domain of the function $(\lambda x: T | P @ E)$ is those values of
$x$ in $T$ for which $P$ holds, and the range is the corresponding
set of values for $E$.
\begin{axdef}
  \_ \comp \_: Exec \cross Exec \fun Exec \\
\where
   (e_1 \comp e_2) =
      (\lambda s:\dom e_1 |
         e_1(s) \in \dom e_2 @
         e_2(e_1(s)))
\end{axdef}
If $\cexec c_1$ maps state $s$ to $s'$
and $\cexec c_2$ maps $s'$ to $s''$,
then $\cexec(c_1 \sand c_2)$ maps $s$ to $s''$.
If either $s$ is not in the domain of $\cexec c_1$ or
$s'$ is not in the domain of $\cexec c_2$,
then $s$ is not in the domain of $\cexec(c_1 \sand c_2)$.
The command $\Skip$ is the unit of sequential conjunction,
and both $\abort$ and $\fail$ are left zeroes.


\subsection{Quantifiers}
\label{sec-cexec-quant}

For a variable $V$ and a state $s$, we define the state `$\unbind~V~s$'
as one whose bindings match those of 
$s$ in every place except $V$, which is completely unconstrained.
\label{def-unbind}
\begin{axdef}
  \unbind: Var \fun State \fun State 
\where
  \unbind~V~s = \{ b:s; x:Val 
                  @ b \oplus \{ V \mapsto x \} \} 
\end{axdef}

Execution of an existentially quantified command $(\exists V @ c)$ from
an initial state $s$ is defined if executing $c$ is defined in the
state $s'$, which is the same as $s$ except that $V$ is unbound.
The resultant state after executing $c$ consists of all
those bindings $b$ in $s$ such that there is a value, $x$,
for $V$ such that execution of $c$ retains the binding
$b \oplus \{ V \mapsto x \}$.
We thus make the following definition of the existential quantifier for
executions.
\label{def-eex}
\begin{axdef}
  \eex: Var \fun Exec \fun Exec
\where
  \eex~V~e = (\lambda s:State | \unbind~V~s \in \dom e \\
\t1  @ \{ b: s | (\exists x:Val @ e(\{b \oplus \{V \mapsto x\} \}) \neq \emptyset)\})
\end{axdef}
Universal quantification behaves in a similar fashion,
except that for $(\eall~V~e)$
to retain a binding $b$, execution of $e$ must retain
$b \oplus \{ V \mapsto x \}$ for all values $x$.
\label{def-eall}
\begin{axdef}
  \eall: Var \fun Exec \fun Exec
\where
  \eall~V~e = (\lambda s:State | \unbind~V~s \in \dom e \\
\t1  @ \{ b: s | (\forall x:Val @ e(\{b \oplus \{V \mapsto x\} \}) \neq \emptyset)\})
\end{axdef}

\section{Refinement}
\label{sec:refinement}
An execution $e_1$ is refined by an execution $e_2$ if and only if
$e_2$ is defined wherever $e_1$ is and they agree on their
outputs whenever both are defined.
This is the usual ``definedness'' order on
partial functions, as used,
for example, by Manna~\cite{MANNA:MTC}: it is simply defined by the
subset relation of functions viewed as sets of pairs.
We define refinement, $\refsto$, as a relation ($\rel$) between $Exec$s.
\label{def-erefsto}
\begin{axdef}
  \_ \erefsto \_: Exec \rel Exec
\where
  e_1 \erefsto e_2 \iff e_1 \subseteq e_2
\end{axdef}

For the commands $Cmd$ in our language, we define refinement in terms
of refinement of the corresponding executions.
\label{def-crefsto}
\begin{axdef}
  \_ \crefsto \_: Cmd \rel Cmd
\where
  c_1 \crefsto c_2 \iff \cexec~c_1 \erefsto \cexec~c_2
\end{axdef}

Finally, refinement equivalence ($\refeq$) is defined for
$Cmd$ and $Exec$ as refinement in both directions.
\label{def-refeq}
\begin{axdef}
  \_ \refeq \_ : Cmd \rel Cmd \\
  \_ \refeq \_ : Exec \rel Exec
\where
  c_1 \refeq c_2 \iff c_1 \crefsto c_2 \land c_2 \crefsto c_1 \\
  e_1 \refeq e_2 \iff e_1 \erefsto e_2 \land e_2 \erefsto e_1 \\
\end{axdef}
Refinement is a preorder --- a reflexive and transitive relation ---
because subset is a preorder on sets.

\subsection{Lattice properties}
\label{sec-lattice}
The refinement relation forms a meet semi-lattice over $Exec$.
\begin{itemize}
  \item `$\erefsto$' is a partial order
  because `$\subseteq$' is a partial order on sets;
  \item meets exist:
     \(
        e_1 \sqcap e_2 = e_1 \cap e_2
     \);
  \item there is a unique bottom element, corresponding to the command
        $\abort$ (recall that
         $\cexec~\abort = \{ \emptyset \mapsto \emptyset \}$).
\end{itemize}

Note that joins do not exist in general,
because $e_1 \cup e_2$ may not be a function;
and there is no top element.
For $e_1 \sqcup e_2$ to be defined, we require that $e_1 \cup e_2$ is a
function and not simply a relation (i.e., $e_1$ returns the same
state as $e_2$ for the states where both are defined).
Even under that condition, the result is not simply $e_1 \cup e_2$,
because that would violate condition (1) of executions.
Instead, we define $e_1 \sqcup e_2$ by adding to $e_1 \cup e_2$
mappings for all states that consist of bindings for which
either $e_1$ or $e_2$ is defined.

\label{def-join}
\label{def-meet}
\begin{axdef}
  \_ \sqcap \_ : Exec \cross Exec \fun Exec \\
  \_ \sqcup \_ : Exec \cross Exec \pfun Exec
\where
  e_1 \sqcap e_2 = e_1 \cap e_2 \\
  (e_1,e_2) \in \dom(\_\sqcup\_) \iff
         (\forall s: \dom e_1 \cap \dom e_2 @ e_1(s) = e_2(s)) \\
  (e_1,e_2) \in \dom(\_\sqcup\_) \implies \\
        \t1 e_1 \sqcup e_2 = 
                (\lambda s: \power (\stateuni (\dom e_1 \cup \dom e_2)) 
                @ \{ b: s | (e_1 \cup e_2)(\{ b \}) \neq \emptyset \})
%
%
\end{axdef}
In \cite{tr00-30}, we show that $\sqcap$ and $\sqcup$
(when the latter is well-defined) preserve the healthiness properties
for executions.

If $e_1 \erefsto e_2$ then their join $e_1 \sqcup e_2$
is defined (and is $e_2$).
As a result, we can define joins for chains.
We first define a chain of executions:
\label{def-chain}
\begin{zed}
  Chain == \{ ec: \nat \fun Exec |
              (\forall i: \nat @ ec(i) \erefsto ec(i+1)) \}
\end{zed}
Note that we define every chain as an infinite sequence; a finite chain is
simply modelled as an infinite chain in which the last element is repeated
infinitely often.
If $ec$ is a chain, then $ec(i) \sqcup ec(i+1)$ exists for all $i$
and is equal to $ec(i+1)$.
We therefore define the join of a chain as follows.
\label{def-chain-join}
\begin{axdef}
  \bigsqcup: Chain \fun Exec
\where
  \bigsqcup ec =  \stateuni(\ran ec)
\end{axdef}

\section{Procedures and parameters}
\label{sec:procedures}
To simplify the semantics, we treat procedures, parameters and
recursion as separate, though related,
concerns in a manner similar to Morgan~\shortcite{MORG:PROCS-PARMS-ABSTR}.

\subsection{Parametrised commands}
\label{def-pcmd}
To deal with parameters, we introduce the notion of a {\em parametrised
command}.
Given a variable $V$ and a command $c$, the expression $\pcmd(V,c)$
denotes the parametrised command ($PCmd$) that,
when provided a term argument $t$, behaves like $c[t/V]$,
i.e., the command $c$ with every occurrence of $V$ replaced by $t$.
In order for
$\pcmd(V,c)$ to be well-formed, we require that $c$ has no free
variables other than $V$, i.e., \mbox{$\free(c) \subseteq \{ V \}$};
any other variables in $c$ must be explicitly quantified.

The semantics of parametrised commands is given by parametrised
executions, which are functions mapping actual parameter terms
to executions.
\label{def-pexec}
\begin{zed}
  PExec == Term \fun Exec
\end{zed}

\subsection{Environments}
To handle procedure definitions and recursion, we
introduce a given set of {\em procedure identifiers\/} ($PIdent$) and
an {\em environment}, which maps procedure identifiers to their corresponding
procedure executions.
\label{def-pident}
\label{def-env}
\begin{zed}
  Env == PIdent \pfun PExec
\end{zed}
Hence we change the definition of $\cexec$ to add an
environment parameter.
\label{def-cexecenv}
\begin{axdef}
  \cexecenv: Env \fun Cmd \fun Exec
\end{axdef}

The definitions we have given in Section~\ref{SEC:execs} do not depend
directly on the environment.
The only change required is to add the environment parameter to the calls on
$\cexec$ for subcomponents, e.g., for an environment $\rho$:
\begin{zed}
  \cexecenv(\rho)(c_1 \pand c_2) =
  (\cexecenv~\rho~c_2) \lcap (\cexecenv~\rho~c_2)
\end{zed}


\subsection{Parametrised executions}

For an actual parameter term $t$,
the execution of a parametrised command $\pcmd(V,c)$ is a function
that is defined for all states $s$
for which evaluation of $t$ is defined for all bindings in $s$,
and for which $c$ is defined when $V$ is bound to the value
of $t$ for each binding in $s$.
We determine the result of executing the parametrised command
by determining its result for each binding $b$ in $s$.
A binding $b$ from a state $s$ will be retained by the execution of
the parametrised command applied to term $t$ if the execution
of $c$ from a state consisting of $\{b\}$, but with the formal parameter
$V$ replaced with the value of term $t$ in binding $b$, retains that
binding.
Recall that the result of executing a command on a singleton
set is either that singleton set or the empty set.
\label{def-pexecenv1}
\begin{axdef}
  \pexecenv: Env \fun PCmd \fun PExec
\where
  \pexecenv(\rho)(\pcmd(V,c)) = (\lambda t:Term @ \\
\t1  (\lambda s: \defined~t | \assign~V~t~s \in \dom(\cexecenv~\rho~c) @ \\
\t2      \{ b: s | (\cexecenv~\rho~c)(\assign~V~t~\{b\}) \neq \emptyset \}))
\end{axdef}
This definition only applies to non-recursive parameterised commands.
The definition of $\pexecenv$ for recursion is presented in
Section~\ref{sec:recursion}.

\subsection{Refinement}
We define refinement between parametrised executions $p_1$ and $p_2$
by requiring refinement for every possible value of the parameter.  We
also define refinement equivalence between parametrised executions in
the obvious way.
\label{def-prefsto}
\begin{axdef}
  \_ \prefsto \_: PExec \rel PExec \\
  \_ \prefeq \_: PExec \rel PExec
\where
  (p_1 \prefsto p_2) \iff
    (\forall t:Term @ (p_1~t) \erefsto (p_2~t)) \\
  (p_1 \prefeq p_2) \iff
    (\forall t:Term @ (p_1~t) \refeq (p_2~t))
\end{axdef}

Refinement between parametrised commands $pc_1$ and $pc_2$ is defined,
as expected, in terms of refinement between the corresponding
parametrised executions.
Because the meaning of a parametrised command depends on the
environment,
this becomes a parameter to the refinement relation;
this parameter is written as a subscript $\rho$.
\label{def-pcrefsto}
\begin{axdef}
  \_ \pcrefsto_{\rho} \_: Env \fun (PCmd \rel PCmd) \\
  \_ \pcrefeq_{\rho} \_: Env \fun (PCmd \rel PCmd)
\where
  (pc_1 \pcrefsto_{\rho} pc_2) \iff
     (\pexecenv~\rho~pc_1 \prefsto \pexecenv~\rho~pc_2) \\
  (pc_1 \prefeq_{\rho} pc_2) \iff
     (\pexecenv~\rho~pc_1 \pcrefeq \pexecenv~\rho~pc_2)
\end{axdef}
When the environment $\rho$ is clear from the context
it may be elided.

\subsection{Procedure call}
\label{sec:procedure-calls}
\label{def-proc-call}

A parametrised command may be directly applied to a term;
the result is a command, the semantics of which is defined as follows.
\begin{zed}
  \cexecenv~\rho~(\pcall(({\pcmd(v,c)}),t)) = \pexecenv~\rho~(\pcmd(v,c))~t
\end{zed}
The syntax of a procedure call is $id(t)$, where $t$ is a term and
$id$ is a procedure identifier.
The parametrised command which is the definition of $id$
in the environment is applied to $t$.
If $id$ is not defined in the environment,
the result of a call on $id$ is $\abort$.
\begin{zed}
  \cexecenv~\rho~\call(id,t) =
    \IF id \in \dom \rho \THEN (\rho~id~t) \ELSE (\cexecenv~\rho~\abort)
\end{zed}

\subsection{Lattice properties}

The lattice properties of $Exec$ can be lifted to $PExec$.
\label{def-lsqcap}
\label{def-lsqcup}
\begin{axdef}
  \_\lsqcap\_ : PExec \cross PExec \fun PExec \\
  \_\lsqcup\_ : PExec \cross PExec \pfun PExec 
\where
  p_1 \lsqcap p_2 = (\lambda t: Term @ p_1~t \sqcap p_2~t) \\
  (p_1,p_2) \in \dom(\_\lsqcup\_) \iff
    (\forall t: Term @ (p_1~t, p_2~t) \in \dom(\_\sqcup\_)) \\
  (p_1,p_2) \in \dom(\_\lsqcup\_) \implies
    p_1 \lsqcup p_2 = (\lambda t: Term @ p_1~t \sqcup p_2~t)
\end{axdef}

\label{def-pchain}
\begin{zed}
  PChain == \{ p: \nat \fun PExec |
               (\forall i: \nat @ p(i) \prefsto p(i+1)) \}
\end{zed}

\label{def-lbigsqcup}
\begin{axdef}
  \lbigsqcup: PChain \fun PExec
\where
  \lbigsqcup~p =
    (\lambda t: Term @ \bigsqcup (\lambda i: \nat @ p~i~t ) )
\end{axdef}

\section{Recursion}
\label{sec:recursion}
If $id$ is an identifier and $pc$ is a parametrised
command, possibly containing instances of $id$, the recursion block
$\rec(id,{pc})$ is also a parametrised command.
Intuitively, a call on the parametric recursion block
$\rec(id,{\pcmd(V,{\ldots id(t) \ldots})})$ is similar to a
call on the Prolog procedure defined by
\verb!id(V) :- ... id(t) ...!.

\subsection{Semantics of recursion blocks}
\label{sec-fixpoints}
A recursion block embeds one or more recursive calls on the block
inside a context.  Thus, a context is a function from one parametrised
command (the recursive call) to another (the entire body of the
recursion block):
\label{def-ctx}
\begin{zed}
  Ctx == PExec \fun PExec
\end{zed}

Intuitively, to represent $\rec(id,\pcmd(V,{\ldots id(t) \ldots}))$,
we define the context
$\C$ such that $\C (p ) = \pcmd(V,{\ldots p(t) \ldots})$.
In this example, $\C$ embeds a call on $p$ in the context given by
$id$, and also provides $p$ with the parameter $t$.
Formally, we define a function that extracts a context from a recursion block.
\label{def-context}
\begin{axdef}
  \context: Env \fun PCmd \pfun Ctx
\where
  \context(\rho)(\rec(id,pc)) = 
  (\lambda p:PExec @ \pexecenv(\rho \oplus \{ id \mapsto p \})(pc))
\end{axdef}
This function is partial because it is only defined for $PCmd$s that
are recursion blocks.

To define the semantics of recursion blocks we use a fixed point
construction, which is defined for all monotonic contexts --- see
Knaster-Tarski Theorem~\cite{NELS:GENERALIZATION}.
We first define the set of monotonic contexts.
\begin{zed}
  MCtx == \{ \C:Ctx | (\forall p,p':PExec @
                (p \prefsto p') \implies (\C(p) \prefsto \C(p')))
             \}
\end{zed}
This monotonicity property holds for every context $\C$ that can be
constructed in our language \cite{tr00-30}.

Now the meaning of a recursion block is the least fixed point of
the context $\C$ provided by the recursion block (written $\fix~\C$),
where:
\label{def-fix}
\begin{axdef}
  \fix: MCtx \fun PExec
\where
  (\forall \C:MCtx @ \fix~\C = \C(\fix~\C))
\also
  (\forall \C:MCtx; p:PExec @ (\C(p) = p) \implies (\fix~\C \prefsto p))
\end{axdef}

Hence the meaning of a recursion block is the least fixed point of
the context corresponding to the recursion block.
\label{def-pexecenv2}
\begin{zed}
  \pexecenv(\rho)(\rec(id,pc)) =
    \fix(\context(\rho)(\rec(id,pc)))
\end{zed}

\subsection{Constructing the fixed point}
\label{sec-fixpoint-construction}
To simplify this and the next section, we use the syntax of
parametrised commands to stand for their $PExec$s and we assume
a fixed environment $\rho$, which is augmented with a single recursive
definition.

The least defined command is $\abort$.
The least defined parametrised command is a parametrised
command with $\abort$ as its body:
\[
   \pabort == \pcmd(V,\abort)
\]
For a recursion based on a monotonic context $\C$, we construct the
chain of programs:
\begin{axdef}
  pc : PChain
\where
  pc = (\lambda i:\nat @ \C\bsup i \esup(\pabort))
\end{axdef}
That is, we have
\begin{zed}
  pc(0) = \C\bsup 0 \esup(\pabort) = \pabort \\
  pc(i+1) = \C\bsup i+1 \esup(\pabort) = \C(pc(i))
\end{zed}

The sequence $pc$ forms a chain
ordered by $\prefsto$ and has a join ($\lbigsqcup pc$).
By the Limit Theorem~\cite{NELS:GENERALIZATION}, $\lbigsqcup pc = \fix~\C$,
as long as $\C$ is a chain-continuous function.
  For a chain $pc$ 
  for which the join $\lbigsqcup pc$ exists,
  a function $\C$ is {\em chain-continuous\/} provided 
  $\C(\lbigsqcup pc) = \lbigsqcup (\lambda i: \nat @ \C(pc(i)))$.
All the contexts $\C$ that can be constructed in our language
are chain-continuous \cite{tr00-30}.

For example, the following equivalences hold in our semantics.
\begin{eqnarray*}
  \rec(p,{\pcmd(X,p(X))}) & \refeq & \abort_1 \\
  \rec(p,{\pcmd(X,{\Spec{X=1}\sand p(X)})}) & \refeq & \pcmd(X,{\Spec{X = 1}\sand \abort}) \\
  \rec(p,{\pcmd(X,{\Spec{X=1}\pand p(X)})}) & \refeq & \abort_1 \\
  \rec(p,{\pcmd(X,{\Spec{X=1}\por  p(X)})}) & \refeq & \abort_1
\end{eqnarray*}
The first is the trival non-terminating recursion.
The second fails if $X$ is bound to a value other than one,
otherwise it is a non-terminating recursion.
The last two both use parallel operators that evaluate both arguments;
hence they both degenerate to a non-terminating recursion.

\subsection{Mutual recursion}

In this paper we do not explicitly handle the definition of a set of
mutually recursive procedures.
Such a set can always be encoded as a single procedure and hence given
a semantics via this encoding.
For example, a set of mutually recursive procedures
\[
        p_1 \defs \pcmd(V_1,C_1) \\
        \vdots \\
        p_n \defs \pcmd(V_n,C_n)
\]
may be encoded as a single procedure
\[
        p \defs \pcmd({(I,V_1,\ldots,V_n) \;},
                 \Spec{I = 1}, C_1 \lor 
                 \ldots \lor
                 \Spec{I = n}, C_n)
\]
where the parameter $I$ (a fresh name) encodes which of the original
procedures is being called and the parameter names $V_1,\ldots,V_n$
are assumed to be distinct.  A call of the form $p_1(t)$ is then
encoded as $p(1,t, \_, \ldots, \_)$.

\section{Refinement laws}
\label{sec:reflaws}

This section presents a number of refinement laws for the step-wise
refinement of logic programs.  Each of these laws has been proven with
respect to the semantics, using the execution properties listed in
Section~\ref{sec:execution:properties}.
Appendix~\ref{sec:reflaws:listing} presents
a summary of refinement laws used in this paper.

\subsection{Algebraic laws}

The commands in the wide-spectrum language obey a number of algebraic
properties. Parallel conjunction and disjunction are
commutative, but sequential conjunction is not. Parallel conjunction
and disjunction as well as sequential conjunction are associative.
Parallel conjunction can be refined to sequential conjunction using the
rule \lref{\pandtosand} because in the sequential form $c_2$ can
assume the context established by $c_1$,
but in the parallel form it cannot:
\[ c_1 \pand c_2 \refsto c_1 \sand c_2 \]
To prove such a law we show 
\[
  \cexec(\rho)(c_1 \pand c_2) \crefsto \cexec(\rho)(c_1 \sand c_2)
 \Equiv
  \cexec~\rho~c_1 \lcap \cexec~\rho~c_2 \subseteq
  \cexec~\rho~c_1 \semi \cexec~\rho~c_2
\]
which can be shown via the definitions of $\lcap$ and $\semi$,
and the properties of executions.

Parallel conjunction and disjunction distribute over each other. Sequential
conjunction distributes over parallel conjunction and disjunction in the
following ways:
\begin{eqnarray*}
  c_1 \sand (c_2 \pand  c_3) & \refeq & (c_1 \sand c_2) \pand  (c_1 \sand c_3)\\
  c_1 \sand (c_2 \por  c_3) & \refeq & (c_1 \sand c_2) \por  (c_1 \sand c_3)\\
   c_1 \pand (c_2 \sand c_3) & \refsto & (c_1 \pand c_2) \sand c_3\\
  (c_1 \por c_2) \sand c_3 & \refeq & (c_1 \sand c_3) \por (c_2 \sand c_3) 
\end{eqnarray*}
The existential and universal quantifiers distribute over 
disjunction and (parallel) conjunction respectively.
\begin{eqnarray*}
  (\exists V @ c_1 \por c_2) & \refeq &
       (\exists V @ c_1) \por (\exists V  @ c_2) \\
  (\forall V @ c_1 \pand c_2) & \refeq &
       (\forall V @ c_1) \pand (\forall V  @ c_2)
\end{eqnarray*}






\subsection{Monotonicity laws}

Each of the language constructs is monotonic with respect to the
refinement relation.  
Monotonicity guarantees that the result of replacing a component
of a program by its refinement is itself a refinement of the original
program.

For  parallel  and sequential conjunction,
and disjunction, there are laws stating that the command
can be refined by refining either of the arguments of the top-level operator;
e.g., the rule \lref{\pormono}:
\[ \Rule{c_1 \refsto c_2; c_3 \refsto c_4}{c_1 \por c_3 \refsto c_2 \por c_4} \]

For the existential and universal quantifiers we give monotonicity laws
stating that the command can be refined by refining the quantified
sub-command; e.g., the rule \lref{\existsmono} for existential
quantifiers:
\[ \Rule{c_1 \refsto c_2}{\exists V @  c_1 \refsto \exists V @ c_2} \]

\subsection{Predicate lifting}

The predicate lifting laws state that predicate operators inside
of specifications can be lifted to their corresponding command
operator. For example, the rule \lref{\liftpand} states that
predicate conjunction within a specification can be lifted to
parallel conjunction at the command level:
\[ \LPSpec{P \land Q} \refeq \LPSpec{P} \pand \LPSpec{Q} \]
Similar rules are given that lift predicate disjunction to command
disjunction, and predicate quantifiers within specifications to the
corresponding command quantifiers.

\subsection{Specification and assumption laws}

Programs can be refined by weakening, removing and introducing
assumptions.  For example, the law \lref{\weakenassumpt} states that
an assumption can be refined by weakening the predicate:
\[ \Rule{A \entails B}{\Assert{A} \refsto \Assert{B}} \]
where for predicates $A$ and $B$, $A$ entails $B$ ($A \entails B$)
iff for all bindings $b$, $A$ is true at $b$ implies $B$ is true
at $b$, i.e., $A \entails B \iff \pred{A} \subseteq \pred{B}$.

A program can be refined
by replacing a specification by an equivalent one; the law
\lref{\equivspec} states that if $P$ and $Q$
are equivalent,  then the program 
$\Spec{P}$ is equivalent to the program $\Spec{Q}$; i.e.,
\[ \Rule{P \equiv Q}{\Spec{P} \refeq \Spec{Q}} \]
where predicates $P$ and $Q$ are said to be equivalent ($P \equiv Q$)
iff for all bindings $b$, $P$ is true at $b$ iff $Q$ is true at $b$,
i.e., $P \equiv Q \iff \pred{P} = \pred{Q}$.

\subsection{Context}

We have already seen that commands are refined in the context of a
program environment $\rho$. A second kind of context is an
``assumption'' context (based on the precondition context used in the
program window inference presented in
\cite{nickhayes-context}). For example, consider the command
`$\Assert{A} \sand S$'; in refining $S$ and its subcommands, we may
assume that the predicate $A$ holds. Rather than propagating the
assumption throughout $S$ and its subcommands, we add the
assumption $A$ to the context of the refinement.

To allow refinement in the context of assumptions, we need to introduce
the notion of pointwise refinement, $\refstopoint$, defined as:
\begin{axdef}
\_ \refstopoint_{\rho} \_, \_ \refeqpoint \_ : Env \fun Cmd \cross Cmd \tfun (Bnd \tfun boolean)
\ST
c_1 \refstopoint_{\rho} c_2 = (\lambda b:Bnd @ \{ b \} \in \dom \cexec~\rho~c_1
\imp \\
	\t2 (\{ b \} \in \dom \cexec~\rho~c_2 \land 
	\cexec~ \rho~c_1~\{ b \} = \cexec~ \rho~ c_2~\{ b \})
\also
c_2 \refeqpoint_{\rho} c_2 = c_1 \refstopoint_{\rho} c_2 \land c_2 \refstopoint_{\rho} c_1
\end{axdef}
As with command refinement we shall omit the environment when it is
clear from the context.
Note that when there are no assumptions (or equivalently the assumption
is $\true$), pointwise refinement is equivalent to command refinement, i.e.,
\begin{zed}
c_1 \refsto c_2 \iff (\true \entails (c_1 \refstopoint c_2))
\end{zed}
The laws \lref{\weakenassumpt} and \lref{\equivspec} can be restated
to include the current context, $\Gamma$.
\begin{sidebyside}
\[ \Rule{\Gamma \entails (A \implies B)}{\Gamma \entails (\Assert{A} \refstopoint
\Assert{B})} \]
\nextside
\[ \Rule{\Gamma \entails (P \iff Q)}{\Gamma \entails (\Spec{P} \refeqpoint
\Spec{Q})} \]
\end{sidebyside}

When refining $c_1$ in the context
$\Assert{A} \sand c_1$ we may assume $A$ holds by extending the
current context; this is encapsulated in the
rule \lref{\assumptincontext}:
\[ \Rule{\Gamma \land A \entails (c_1 \refstopoint c_2)}{\Gamma \entails \Assert{A} \sand c_1 \refstopoint \Assert{A} \sand c_2}
\]
Similarly for specifications we have the rule \lref{\specincontext}:
\[ \Rule{\Gamma \land P \entails (c_1 \refstopoint c_2)}{\Gamma \entails \Spec{P} \sand c_1 \refstopoint \Spec{P} \sand c_2}
\]

If a command $c_1$ refines to $c_2$ in a context $\Gamma$, 
then it refines to $c_2$ in any stronger context.
In particular, given a refinement law of the form $c_1 \refsto c_2$, then
$c_1 \refstopoint c_2$ in any context, i.e.,
\[ \Rule{c_1 \refsto c_2}{\Gamma \entails (c_1 \refstopoint c_2)}\]
This means that any of the (command) refinement laws 
that do not have proof obligations,
such as the algebraic and predicate laws given earlier,
can be applied in any context. 
We also observe that each of our language constructs are
monotonic with respect to the pointwise refinement relation in any
context. For example, for disjunction the following law holds.
\[ \Rule{\Gamma \entails (c_1 \refstopoint c_2); \Gamma \entails (c_3 \refstopoint c_4)}{\Gamma \entails (c_1 \por c_3 \refstopoint c_2 \por c_4)} \]
Notice this law is more general than \lref{\pormono}, which is an instance
of the above law with the assumption context being $\true$.

\subsection{Recursion introduction law}

\def \refstoenv {\refstopoint_{\rho}}
\def \refstoenvp {\refstopoint_{\rho'}}
\def \refeqenv {\refeqpoint_{\rho'}}

We desire a refinement law that allows us to refine a non-recursive
parameterised command into a recursive block.
That is, 
for $pc$ a parameterised command and 
$\C(id)$ a command that may depend on
the (fresh) procedure identifier
$id$ and the variable $V$,
we want to derive a law with
the following
as the conclusion.
\[
	pc \refstoenv \re id @ V \prm \C(id) \er
\]

We will derive such a law using well-founded induction, the general form of
which is, for some predicate $\phi$ and well-founded relation $\prec$,
\[
	\Rule{(\all Y @ Y \prec V \imp \phi(Y)) \entails \phi(V)}
		{\all X @ \phi(X)}
\]
We take $\phi(X)$ to be
$pc(X) \refstoenv (\re id @ V \prm \C(id) \er)(X)$.
Given this, and the definition of 
refinement for parameterised commands,
the predicate $\all X @ \phi(X)$ on the bottom line 
is equivalent to our initially stated refinement,
\[
	pc \refstoenv \re id @ V \prm \C(id) \er
\]
Now we instantiate the top line of the well-founded induction law using the
same~$\phi$.
\[
	(\all Y:Term @ Y \prec V \imp 
		pc(Y) \refstoenv (\re id @ V \prm \C(id) \er)(Y)) \entails \\
		\t1 pc(V) \refstoenv (\re id @ V \prm \C(id) \er)(V) 
\]
As $id$ was a fresh name, we define 
$\rho' = \rho \union \{id \mapsto \re id @ V \prm \C(id) \er\}$.
The top line can now be written as
\[
	(\all Y:Term @ Y \prec V \imp pc(Y) \refstoenvp id(Y)) \entails \\
		\t1 pc(V) \refstoenvp (\re id @ V \prm \C(id) \er)(V) 
\]
Now we simplify the expression on the right-hand side of the entailment
so as to make the whole expression useful as a proof obligation
for the recursion introduction law.
\begin{align*}
	~& pc(V) \refstoenvp (\re id @ V \prm \C( id) \er)(V) \\
	 \equiv~& \text{unfold} \\
	~& pc(V) \refstoenvp (V \prm \C( \re id @ V \prm \C( id) \er))(V) \\
	 \equiv~& \text{parameter application; definition of $id$ in $\rho'$} \\
	~& pc(V) \refstoenvp \C(id) 
\end{align*}

Noting that $A \imp c \refstoenv c' \equiv \Ass{A},c \refstoenv c'$,
the process produces the following recursion introduction law,
\lref{\recursionintro}.
\[
	\Rule{(\all Y:Term @ \Ass{Y \prec V} \sand pc(Y) \refstoenvp id(Y)) \entails
		pc(V) \refstoenvp \C(id)}
		{pc \refsto_{\rho} \re id @ V \prm \C(id) \er}
\]
\newcommand{\COMMENT}[1]{}

\COMMENT{
The use of this law may be inelegant, since one must already know the
structure, $\C$, of the body of the recursion block before the refinement law
can be applied.
A more natural way to use the law is to include the 
antecedent of the hypothesis of the law as an assumption inside the body of the
recursive block.
\begin{equation}
\label{rec:law}
  pc \refeqenv \rec(id,{
    \pcmd(V,{\Assert{ \all Y:Term @ \Ass{Y \prec V} \sand pc(Y) \refstoenvp
	id(Y) } \sand pc(V)})})
\end{equation}
In this law, which has no proof obligations,
the parameterised command $pc$ is refined to a recursive block, the body of
which is just $pc(V)$, with an assumption  -- the inductive hypothesis --
that $pc(Y)$ may be refined to a
recursive call for $Y \prec V$.
The refiner then focuses on $pc(V)$ and refines it using the inductive
hypothesis, possibly (probably) introducing recursive calls to $id$, i.e.,
refining to to some command $\C(id)$.
This is exactly the proof obligation for the law given above.
Using (\ref{rec:law})
is more straightforward, however, it requires the syntax of
refinement and commands to be allowed as predicates inside assumptions,
which we wish to avoid.
However for simplicity in our refinements we use (\ref{rec:law}) to introduce
recursion, 
and appeal to the derived law for correctness.
}

\COMMENT{
Suppose $pc$ is a parameterised command, $(\_ \prec \_)$ is a
well-founded relation on terms, and 
${\cal C}$ is some context
that expects a parametrised command
as a parameter and returns a parametrised command.
 To refine $pc$
to a recursion block, the body of which is defined via ${\cal C}$, 
it is sufficient to
show that $pc$ is refined by an instance of the context, where the
parameter is the same as $pc$, but has the additional assumption that
its parameter is less than the original parameter to $pc$.

\[ \Rule{\all x:X @ V \prm \Assert{V = x} \sand pc(V) \refsto {\cal C} (V \prm \Assert{V \prec x}
\sand pc(V))}{pc \refsto \rec(id, {{\cal C}(id)})} \]
This law is similar to the recursion introduction refinement law given by Back
and von Wright for the imperative refinement calculus \cite{Back:98}.

In practice, this application of this law can inelegant due to the size of the
proof obligation, and
that the form of the recursive procedure (i.e.,
the structure of ${\cal C}$) must be pre-determined.
We prefer to refine our command $pc$, with an assumption that any instances
of $pc$ with an actual parameter $y$
less than $V$ may be refined to a recursive
call to $id$.
This is given by the following predicate $IH$ (inductive hypothesis).
\[
IH(V) == \forall y: Term @ \Assert{y \prec V} \sand pc(y)
\refstopoint id(y)
\]

In the law \lref{\recursionintro}, the inductive hypothesis is
directly encoded as an assumption within the recursion block.  
This
makes the inductive hypothesis available in the context when the body
of the recursion block is being refined.  
Since the inductive hypothesis includes pointwise refinement, we need to extend
the predicates allowed inside assumptions 
(but not in specifications) to include pointwise refinement between commands.
The \lref{\recursionintro} law is as follows.
\[
  pc \prefeq \rec(id,{
    \pcmd(V,{\Assert{ IH(V) } \sand pc(V)})})
\]
A proof of this law may be found in \cite{tr00-30}.
}

\subsection{Example}
\label{sec-fac-xmp}

We derive a recursive implementation of a factorial function from
the definition of factorial, which is:
\[
  fact(0) = 1 \\
  fact(n+1) = fact(n) \times (n+1), \;\; {\mbox {\rm for}} \;\; n \geq 0
\]

Starting with an initial abstract specification $S$, we wish to refine
this to a concrete program $C$, such that $C$ can easily be translated
(automatically) into an executable program. We adopt
a step-wise refinement approach, where we introduce a number of 
intermediate programs $I_j$ (where $1 \leq j \leq n$ for some natural $n$),
representing refinements of the previous program, i.e.,
$S \refsto I_1 \refsto \ldots \refsto I_n \refsto C$. This is achieved
by making use of the transitive nature of the refinement relations.

Monotonicity rules are used to transform a subcommand of the program. The
subcommand to be refined is indicated by the markers $\llcorner$ and
$\lrcorner$.  For binary commands we use the more specific
monotonicity rules used to focus on the left or right-hand side 
of a command. Often a single step in the example
actually corresponds to the application of multiple nested
monotonicity rules.

Our factorial program has parameters $U$ and $V$.
It can assume $U \in \nat$ and must establish $V = fact(U)$.
\[
	(U,V) \prm \Assert{U \in \nat} \sand \Spec{V=fact(U)}
\]
We make use of \lref{\recursionintro}, which allows one to assume
\[
	\all U1, V1 @ \Assert{U1 \prec U}, \Assert{U1 \in \nat}, 
		\Spec{V1 = fact(U1)} \refstoenvp f(U1, V1)
\]
where $f$ is a fresh procedure identifier and
$\rho' = \rho \union \{f \mapsto \re f @ (U,V) \prm \C(f) \er\}$
and refine
\[
	\Assert{U \in \nat} \sand \Spec{V=fact(U)}
\]
in this context.
\COMMENT{
\begin{tabbing}
$\prefsto$ \= \kill
\> \ $@$ \=  $(U,V) \prm \Assert{U \in \nat} \sand \Spec{V=fact(U)}$ \+\\
$\refeq$ \> \lref{\recursionintro}:   $<$ is well-founded on $\nat$\+\\
$\re f @ (U,V) \prm {}$ \\
\quad \= $\Assert{\forall U1,V1 @ \Assert{U1<U} \sand
        \Assert{U1 \in \nat} \sand \Spec{V1 = fact(U1)} 
        \refstopoint f(U1,V1)} \sand$ \+\\
$\Assert{U \in \nat} \sand \cout{\Spec{V = fact(U)}}$ \-\\
$\er$
\end{tabbing}
}
Focusing on the specification 
$\Spec{V = fact(U)}$, we can assume the assumptions prior to the
specification; in order that this assumption (and assumptions introduced
to the context later) can be used we now use
the pointwise refinement relation with the extended environment $\rho'$.
\begin{tabbing}
  $\prefsto 1 @$ \= \+ \kill
Assumption 1: \\
  ~~~~$\forall U1,V1 @ \Assert{U1<U} \sand
        \Assert{U1 \in \nat} \sand \Spec{V1 = fact(U1)} \refstoenvp f(U1,V1)$\\
Assumption 2: $U \in \nat$\\
\ $@$ \= $\Spec{V=fact(U)}$ \\
$\refstopoint$ \> case analysis, Assumption 2 \+\\
$(\Spec{U=0} \sand \cout{\Spec{V=fact(U)}}) \por (\Spec{U>0} \sand \Spec{V=fact(U)})$\-\\
$\refstopoint$ \> \=\lref{\specincontext} \+\+\\
Assumption 3: $U = 0$\\
\ $@$ \= $\Spec{V=fact(U)}$\\
$\refstopoint$ \> \lref{\equivspec} ($U=0 \imp (V = fact(U) \iff V =1)$)\+\\
$\Spec{V=1}$ \-\-\-\\
  \> $(\Spec{U=0} \sand \cin{\Spec{V=1}}) \por (\Spec{U>0} \sand \cout{\Spec{V=fact(U)}})$
\end{tabbing}
Focusing on the second occurrence of $\Spec{V = fact(U)}$, we can again
make a contextual assumption, this time from the specification $\Spec{U > 0}$:
\begin{tabbing}
$\prefsto 1 @ 2 @$ \= \+ \kill
Assumption 4: $U > 0$\\
\ $@$ \= $\Spec{V=fact(U)}$\\
$\refstopoint$ \> \lref{\equivspec}, factorial def., assumptions \+\\
 $\Spec{\exists U1,V1 @ U1 = U-1 \land V1 = fact(U1) \land V = V1 * U}$ \-\\
$\refstopoint$ \> \lref{\liftexists}, \lref{\liftpand}, \lref{\pandtosand} (x2)  \+\\
$\exists U1,V1 @ \Spec{U1 = U-1} \sand \Spec{V1 = fact(U1)} \sand
\Spec{V = V1 * U}$ \-\\
$\refstopoint$ \> \lref{\assumptafterspec} (x2) \+\\
$\exists U1,V1 @$ \= $\Spec{U1 = U-1} \sand$ \\
  \> $\cout{\Assert{U1 < U} \sand \Assert{U1 \in \nat} \sand \Spec{V1 = fact(U1)}} \sand$ \\
  \> $\Spec{V = V1 * U}$
\end{tabbing}
Next we use Assumption 1 to introduce a recursive call, making use of the
assumption $U1 = U-1 $ to discharge the variant on the well-founded
relation.

\begin{tabbing}
$\prefsto 1 @ 2 @ 4 @$ \= \+ \kill
\ $@$ \=  $\Assert{U1 < U} \sand \Assert{U1 \in \nat} \sand \Spec{V1 = fact(U1)}$\\
$\refstopoint$ \> Assumption 1 \+\\
$f(U1,V1)$
\end{tabbing}
Putting this all together and removing the assumptions, using 
\lref{\removeassumpt}, we get
\begin{tabbing}
	$\hskip 5mm \refstoenvp$ \=  \kill 
	\> $\Assert{U \in \nat} \sand \Spec{V=fact(U)} $\\
	$\hskip 5mm \refstoenvp$ \> $(\Spec{U=0} \sand \Spec{V=1}) \por$ \\
	\> $(\Spec{U>0} \sand (\exists U1,V1 @ 
		\Spec{U1 = U-1} \sand f(U1,V1) \sand \Spec{V = V1 * U}))$
\end{tabbing}

Closing the introduction of recursion, we have proved:
\begin{tabbing}
$\hskip 5mm \refstoenvp$ \= \kill
	\> $(U,V) \prm \Assert{U \in \nat} \sand \Spec{V=fact(U)}$ \\
$\hskip 5mm \refsto_{\rho}$ \> $\re f @ (U,V) \prm {}$ \\
	\> \quad $(\Spec{U=0} \sand \Spec{V=1}) \por$ \\
	\> \quad $(\Spec{U>0} \sand (\exists U1,V1 @ \Spec{U1 = U-1} \sand f(U1,V1) \sand
	\Spec{V = V1 * U}))$ \\
\> $\er$
\end{tabbing}
To translate this (informally) into Prolog, we:
\begin{itemize}
  \item turn the recursion block into a recursive
        procedure;
  \item implement the arithmetic specifications using \verb|is|;
  \item express the disjunction using separate clauses;
  \item make the existential quantification implicit.
\end{itemize}
The result is:
\begin{isabelle}
  f(U,V) :- U=0, V=1. \\
  f(U,V) :- U>0, U1 is U-1, f(U1,V1), V is V1*U.
\end{isabelle}

\section{Example: N-queens} 
\label{nqueens}

In this section we present a non-trivial logic program
refinement to demonstrate some refinement
techniques that have been developed.  The case study chosen is the 
\nqueens\ problem.  It is a
generalisation of a chess problem, where eight queens are to be placed on
a chess board so that no two queens may attack each other (they do not
share a row, column, or diagonal).  The generalisation is to have $N$
queens on an $N \cross N$ board.

\subsection{Specification of \nqueens}
\label{nq:specification}

We first describe and motivate our specification of the \nqueens\
problem.
Given a natural number $N$,
a solution to the \nqueens\ problem can be represented by
a list $S$ of length $N$ that contains numbers between 1 and $N$.
The row number of the queen in column $i$ is given by $S(i)$,
i.e., a queen at location (4,5) is indicated by
$S(4) = 5$.
This representation
implicitly guarantees that there is only a single queen in each column.

We define a predicate $psoln(S,N)$, which checks that in the list $S$
there are no row clashes---no two row numbers are the same---and
no diagonal clashes---the absolute value of the difference in
the column numbers does not equal the absolute difference in the
row numbers---as well as requiring the elements of $S$ are valid row numbers
in the range 1 to $N$.
\[
  psoln(S,N) == list(S) \land (\all i: 1..\#S @ S(i) \in 1..N \land \\
        \t1 (\all j: i+1..\#S @ 
              S(i) \neq S(j) \land
              \abs{i - j} \neq \abs{S(i) -S(j)}))
\]
Our specification of \nqueens\ constrains the length
of a solution $S$ to be $N$.
\[
        nqueens \defs (N, S) \prm 
                \Ass{N \in \nat}, 
                \Spec{\#S = N \land psoln(S,N)}
\]

\subsection{Refinement of \nqueens}
\label{nq:refinement}

We note the following derived property of $psoln$.
\begin{align}
        ~& psoln(\HT,N) \iff
                \left(\begin{array}{l}
                    psoln(T,N) \land H \in 1..N \land \\
                    notrow(H,T) \land notdiag(H,T)
                \end{array}\right)
                \label{psoln:sub} 
\end{align}
where,
for $H$ a natural number and $T$ a list of natural numbers,
\begin{align}
    ~& notrow(H,T) == (\all i : 1..\#T @ H \neq T(i)) \label{notrow} \\
    ~& notdiag(H,T) == (\all i : 1..\#T @ i \neq \abs{H - T(i)}) \label{notdiag}
\end{align}

We implement the \nqueens\ solution
using an accumulator-style program, by introducing a
partial solution $P$ as a parameter.
$P$ starts as an empty list,
and is extended on each recursive call until a full
solution of length $N$ is built.
The accumulator version of N-queens includes $psoln(P,N)$
as an assumption, and  must establish that the full solution $S$
has a suffix of $P$ and satisfies $psoln(S,N)$.
\[
        nqacc \defs (N, P, S) \prm \Ass{N \in \nat}, 
                \Ass{psoln(P,N)}, \\
                \t1 \Spec{P \suffix S \land \#S = N \land psoln(S,N)}
\]
We implement {\em nqueens} by calling {\em nqacc} with
the empty list, i.e.,
\[ nqueens(N,S) \refeq nqacc(N,\el,S) \]
The equivalence of these programs can be seen to hold by
observing:
\[
        psoln(\el,N) \land \el \suffix S
\]
Our task is now to refine {\em nqacc}.
As with the refinement of the factorial program in
Section~\ref{sec-fac-xmp}, 
we use the \lref{\recursionintro} law.
We may assume the following inductive hypothesis during the refinement.
\begin{equation}
\begin{split}
\label{nq:ind:hyp}
	~& (\all N,P1,S @ \\
	~&  \t1 \Assert{P1 \prec P} \sand \Ass{N \in \nat}, \Ass{psoln(P1,N)}, \\
	~&  \t1 \Spec{P1 \suffix S \land \#S = N \land psoln(S,N)} \\
	~&  \t1 \refstopoint nq(N,P1,S))
\end{split}
\end{equation}
where $P1 \prec P$ is the well-founded relation which holds when
$N \geq \#P1 > \#P$. (Note that $P1$ is longer than $P$, but the
well-foundedness is maintained by the upper bound of $N$.)

Now we begin the refinement of the body of $nqacc$.
We apply
the \lref{\caseanalysis} law,
to split the specification into the cases $\#P = N$ and $\#P < N$,
because $psoln(P,N)$ implies $\#P \leq N$.
\[
	\Ass{N \in \nat}, \Ass{psoln(P,N)}, \\
	(\Spec{\#P = N},
		 \Spec{P \suffix S \land \#S = N \land psoln(S,N)} \\
	\lor \\
	\Spec{\#P < N},
		 \Spec{P \suffix S \land \#S = N \land psoln(S,N)}) \\
\]

The first disjunct is the base case:
\[
        \Spec{\#P = N} , 
        \Spec{P \suffix S \land \#S = N \land psoln(S,N)} 
\]
From the context we know $psoln(P,N)$.
Hence the above is equivalent to
\[
        \Spec{\#P = N} \sand \Spec{S = P}
\]


Next we focus on the branch that requires a recursive
call.  
\[
        \Spec{\#P < N} , 
        \Spec{P \suffix S \land \#S = N \land psoln(S,N)}
\]
Since we are accumulating the solution, we prepend 
an element $X$ to the front of $P$.
We note that 
\[ \#P < \#S \imp (P \suffix S \iff (\exists X @ \XP \suffix S)) \]
Using the law \lref{\specincontext}
we refine to
\[
       \Spec{\#P < N} \sand 
        \Spec{(\exists X @ \XP \suffix S) \land \#S = N \land psoln(S,N)}
\]
The predicate $psoln(S,N)$ has the property that any suffix $P$ of
$S$ also satisfies $psoln(P,N)$.
Hence the above is equivalent to the following.
\[
  \Spec{\#P < N} \sand \\
  \Spec{(\exists X @ psoln(\XP,N) \land \XP \suffix S) \land
         \#S = N \land psoln(S,N)}
\]
We extend the scope of $X$ to the end of the program,
then use \lref{\liftexists} to take it to the program level.
We also convert the conjunction involving $psoln(\XP,N)$ to a
sequential conjunction, with $psoln$ on the left-hand side
(using \lref{\liftpand} and \lref{\pandtosand}).
\[
        \Spec{\#P < N} , \\
        (\exists X @ \Spec{psoln(\XP,N)} \sand 
                        \Spec{\XP \suffix S \land \#S = N \land psoln(S,N)})
\]
We now establish the assumptions needed to match the inductive hypothesis.
Since $N \geq \#\XP > \#P$ we may introduce $\XP \prec P$.
Since $N \in \nat$ is in context we may introduce it as an assumption.
We also introduce $\Ass{psoln(\XP,N)}$ using law \lref{\establishassumpt}.
\[
  \Spec{\#P < N} \sand \\
  (\exists X @ \Spec{psoln(\XP,N)} \sand \\
\t1  \Ass{\XP \prec P} \sand \Ass{N \in \nat} \sand \Ass{psoln(\XP,N)} \sand \\
\t1  \Spec{\XP \suffix S\land \#S = N \land psoln(S,N)})
\]
We may now use the inductive hypothesis
\refeqn{nq:ind:hyp} to introduce a recursive call on
$nq(N,\XP,S)$.
\[
  \Spec{\#P < N} \sand 
  (\exists X @ \Spec{psoln(\XP,N)} \sand nq(N, [X|P], S))
\]
By observing the equivalence
\refeqn{psoln:sub}, and noting that
the context includes $psoln(P,N)$, 
we may simplify $\Spec{psoln(\XP,N)}$ to $\Spec{X \in 1..N \land
notrow(X,P) \land notdiag(X,P)}$ using \lref{\specincontext}.  
We also lift the conjunctions
to the program level using \lref{\liftpand}.
Now closing the introduction of recursion, we get
\[
    \re nq @ (N, P, S) \prm \\
        \t1 \Ass{N \in \nat}, 
        \Ass{psoln(P,N)}, \\
                \t1 (\Spec{\#P = N}, \Spec{P = S} \\
                \t1 \lor \\
                \t1 \Spec{\#P < N} , \\
                \t1 (\exists X @ (\Spec{X \in 1..N} \pand 
                        \Spec{notrow(X,P)} \pand
                        \Spec{notdiag(X,P)}) \sand \\
                        \t2 nq(N, [X|P], S))) \\
    \er
\]
This is the overall structure of the program.  However we must still
implement $\Spec{X \in 1..N}$, $\Spec{notrow(X,P)}$ and 
$\Spec{notdiag(X,P)}$.

\subsection{Subproblems}

The implementation of $\Spec{X \in 1..N}$ via a recursive procedure is
straightforward and is omitted here.

To refine the other two specifications we observe that they each 
require that a property $\P$ is established for every element in a list
(cf. \refeqn{notrow} and \refeqn{notdiag}). However
universal quantification is not executable. To remove the quantification
we recursively establish the property $\P$ for each element
in the list. This is encapsulated in the following law.

\[
\mbox{\sf\propertyoverlist}\\
  \Spec{list(L)} \land \Spec{\all i: 1..\#L @ \P(V,L(i))} \refsto proc(V,L) \\
  \mbox{where} \\
  proc \defs \re p @ (V,L) \prm \\
           \t1 \Spec{L = \el} \lor
               (\exists H,T @ \Spec{L = \HT} \sand \Spec{\P(V,H)} \sand p(V,T))
           \er
\]
We use this law to
derive an implementation of $\Spec{notrow(X,P)}$.
\[
    \Spec{list(P)} \land \Spec{\forall i: 1..\#P @ X \neq P(i)} \refsto norowclash(X,P) \\
  \mbox{where} \\
        norowclash \defs \re norowc @ (X,P) \prm \\
                \t1 \Spec{P = \el} \lor
                    (\exists H,T @ \Spec{P = \HT} \sand
                        \Spec{X \neq H} \sand norowc(X, T)) \er
\]

In a similar way we derive an implementation of $\Spec{notdiag(X,P)}$,
though in this case we use a more general law
\lref{\propertyoverlistindexed}.
\[
  \mbox{\lref{\propertyoverlistindexed}}\\
  \Spec{list(L)} \land \Spec{\all i: 1..\#L @ \P(i,V,L(i))} \refsto proc(V,L,1) \\
  \mbox{where} \\
  proc \defs \re p @ (V,L,J) \prm \\
           \t1 \Spec{L = \el} \lor
             (\exists H,T @ \Spec{L = \HT} \sand \Spec{\P(J,V,H)} \sand p(V,T,J+1))
           \er
\]
The implementation of $\Spec{notdiag(X,P)}$ follows.
\[
  \Spec{list(P)} \land \Spec{\forall i: 1..\#P @ i \neq \abs{X-P(i)}} \refsto nodiagAcc(X,P,1) \\
  \mbox{where} \\
        nodiagAcc \defs \re nod @ (X,P,J) \prm \\
                \t1 (\Spec{P = \el} \lor \\
                \t1 (\exists H,T @ \Spec{P = \HT} \sand 
                        \Spec{J \neq \abs{X - H}} \sand 
                        nod(X,T,J+1))) \er
\]
Whereas earlier laws are primitives corresponding to language
constructs,
the laws \lref{\propertyoverlist} and \lref{\propertyoverlistindexed}
are examples of higher-level refinement laws that encapsulate a design
pattern for a particular data structure, in this case lists.

\subsection{Prolog implementation}

The refinement is completed by converting any parallel conjunctions to
sequential conjunctions using the law \lref{\pandtosand}. From this program
the following Prolog code can be generated by performing the translations
noted in Section~\ref{sec-fac-xmp}.

\begin{isabelle}
\normalsize
\T1     nqueens(N,S) :- nqacc(N,[],S). \\[2ex]
\T1     nqacc(N,P,S) :- length(P,N), S = P. \\
\T1     nqacc(N,P,S) :- \\
\T2         length(P,M), M < N, \\
\T2         memrng(X,N), norowclash(X,P), nodiagAcc(X,P,1), \\
\T2         nqacc(N,[X|P], S). \\[2ex]
\T1     memrng(X,N) :- N > 0, X = N. \\
\T1     memrng(X,N) :- N > 0, Nm1 is N-1, memrng(X,Nm1). \\[2ex]
\T1     norowclash(\_,[]). \\
\T1     norowclash(X, [H|T]) :- X =\verb|\|= H, norowclash(X,T). \\[2ex]
\T1     nodiagAcc(\_,[],\_). \\
\T1     nodiagAcc(X, [H|T],J):- \\
\T2         XmH is X - H, abs(XmH,AbsXmH), \\
\T2         J1 is J+1, AbsXmH =\verb|\|= J, \\
\T2         nodiagAcc(X, T, J1).
\end{isabelle}

\section{Tool support}
\label{sec:tool}
\newcommand{\isastyle}{\tt}
\newcommand{\isatext}[1]{{\isastyle #1}}

To support the refinement calculus, a prototype tool, \reflptool{} 
\cite{Hemer:01}, has
been developed based on the Isabelle theorem prover
\cite{isabelle94}. \reflptool{} includes an embedding of the
wide-spectrum language (and its associated semantics) in Isabelle/HOL.
All of the refinement laws
listed in Appendix~\ref{sec:reflaws:listing} have been
proven using the tool,
and the factorial example presented in Section~\ref{sec:reflaws}
has been performed.


Isabelle is a generic interactive theorem prover, supporting a variety
of different object logics, including first-order logic, higher-order
logic and Zermelo Frankel set theory.     Isabelle/HOL provides the {\tt
  datatype} facility, used to declare recursive data structures.  Isabelle/HOL
also provides the {\tt primrec} construct, allowing the user to
define primitive recursive operators on recursive datatypes.

\subsection{Tool architecture}

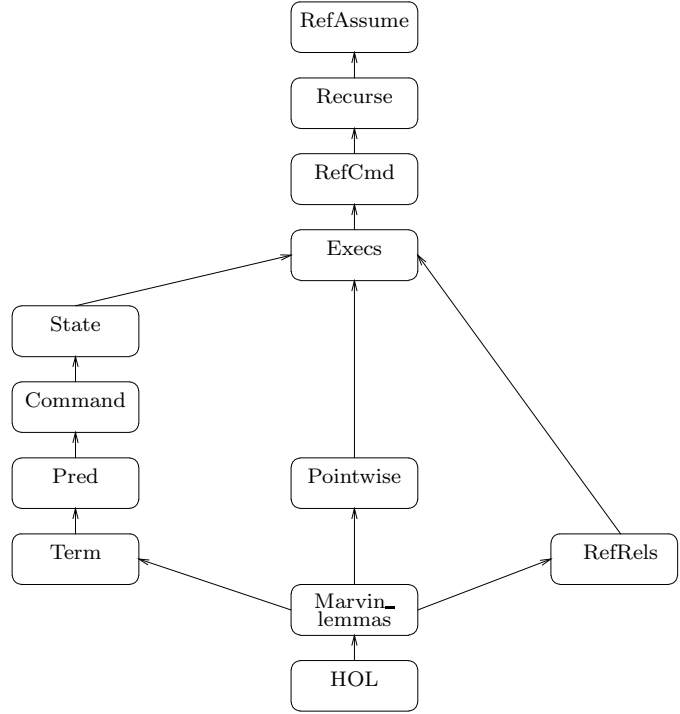
\begin{figure}[htb]
\begin{center}
\setlength{\unitlength}{0.00045in}
\begingroup\makeatletter\ifx\SetFigFont\undefined%
\gdef\SetFigFont#1#2#3#4#5{%
  \reset@font\fontsize{#1}{#2pt}%
  \fontfamily{#3}\fontseries{#4}\fontshape{#5}%
  \selectfont}%
\fi\endgroup%
{
\begin{picture}(7899,8439)(0,-10)
\put(3417,7917){\arc{210}{1.5708}{3.1416}}
\put(3417,8307){\arc{210}{3.1416}{4.7124}}
\put(4707,8307){\arc{210}{4.7124}{6.2832}}
\put(4707,7917){\arc{210}{0}{1.5708}}
\path(3312,7917)(3312,8307)
\path(3417,8412)(4707,8412)
\path(4812,8307)(4812,7917)
\path(4707,7812)(3417,7812)
\put(3417,6117){\arc{210}{1.5708}{3.1416}}
\put(3417,6507){\arc{210}{3.1416}{4.7124}}
\put(4707,6507){\arc{210}{4.7124}{6.2832}}
\put(4707,6117){\arc{210}{0}{1.5708}}
\path(3312,6117)(3312,6507)
\path(3417,6612)(4707,6612)
\path(4812,6507)(4812,6117)
\path(4707,6012)(3417,6012)
\put(3417,7017){\arc{210}{1.5708}{3.1416}}
\put(3417,7407){\arc{210}{3.1416}{4.7124}}
\put(4707,7407){\arc{210}{4.7124}{6.2832}}
\put(4707,7017){\arc{210}{0}{1.5708}}
\path(3312,7017)(3312,7407)
\path(3417,7512)(4707,7512)
\path(4812,7407)(4812,7017)
\path(4707,6912)(3417,6912)
\put(3417,5217){\arc{210}{1.5708}{3.1416}}
\put(3417,5607){\arc{210}{3.1416}{4.7124}}
\put(4707,5607){\arc{210}{4.7124}{6.2832}}
\put(4707,5217){\arc{210}{0}{1.5708}}
\path(3312,5217)(3312,5607)
\path(3417,5712)(4707,5712)
\path(4812,5607)(4812,5217)
\path(4707,5112)(3417,5112)
\put(117,4317){\arc{210}{1.5708}{3.1416}}
\put(117,4707){\arc{210}{3.1416}{4.7124}}
\put(1407,4707){\arc{210}{4.7124}{6.2832}}
\put(1407,4317){\arc{210}{0}{1.5708}}
\path(12,4317)(12,4707)
\path(117,4812)(1407,4812)
\path(1512,4707)(1512,4317)
\path(1407,4212)(117,4212)
\put(117,3417){\arc{210}{1.5708}{3.1416}}
\put(117,3807){\arc{210}{3.1416}{4.7124}}
\put(1407,3807){\arc{210}{4.7124}{6.2832}}
\put(1407,3417){\arc{210}{0}{1.5708}}
\path(12,3417)(12,3807)
\path(117,3912)(1407,3912)
\path(1512,3807)(1512,3417)
\path(1407,3312)(117,3312)
\put(117,2517){\arc{210}{1.5708}{3.1416}}
\put(117,2907){\arc{210}{3.1416}{4.7124}}
\put(1407,2907){\arc{210}{4.7124}{6.2832}}
\put(1407,2517){\arc{210}{0}{1.5708}}
\path(12,2517)(12,2907)
\path(117,3012)(1407,3012)
\path(1512,2907)(1512,2517)
\path(1407,2412)(117,2412)
\put(117,1617){\arc{210}{1.5708}{3.1416}}
\put(117,2007){\arc{210}{3.1416}{4.7124}}
\put(1407,2007){\arc{210}{4.7124}{6.2832}}
\put(1407,1617){\arc{210}{0}{1.5708}}
\path(12,1617)(12,2007)
\path(117,2112)(1407,2112)
\path(1512,2007)(1512,1617)
\path(1407,1512)(117,1512)
\put(3417,1017){\arc{210}{1.5708}{3.1416}}
\put(3417,1407){\arc{210}{3.1416}{4.7124}}
\put(4707,1407){\arc{210}{4.7124}{6.2832}}
\put(4707,1017){\arc{210}{0}{1.5708}}
\path(3312,1017)(3312,1407)
\path(3417,1512)(4707,1512)
\path(4812,1407)(4812,1017)
\path(4707,912)(3417,912)
\put(3417,117){\arc{210}{1.5708}{3.1416}}
\put(3417,507){\arc{210}{3.1416}{4.7124}}
\put(4707,507){\arc{210}{4.7124}{6.2832}}
\put(4707,117){\arc{210}{0}{1.5708}}
\path(3312,117)(3312,507)
\path(3417,612)(4707,612)
\path(4812,507)(4812,117)
\path(4707,12)(3417,12)
\put(3417,2517){\arc{210}{1.5708}{3.1416}}
\put(3417,2907){\arc{210}{3.1416}{4.7124}}
\put(4707,2907){\arc{210}{4.7124}{6.2832}}
\put(4707,2517){\arc{210}{0}{1.5708}}
\path(3312,2517)(3312,2907)
\path(3417,3012)(4707,3012)
\path(4812,2907)(4812,2517)
\path(4707,2412)(3417,2412)
\put(6492,1617){\arc{210}{1.5708}{3.1416}}
\put(6492,2007){\arc{210}{3.1416}{4.7124}}
\put(7782,2007){\arc{210}{4.7124}{6.2832}}
\put(7782,1617){\arc{210}{0}{1.5708}}
\path(6387,1617)(6387,2007)
\path(6492,2112)(7782,2112)
\path(7887,2007)(7887,1617)
\path(7782,1512)(6492,1512)
\path(4062,612)(4062,912)
\path(4092.000,792.000)(4062.000,912.000)(4032.000,792.000)
\path(4062,1512)(4062,2412)
\path(4092.000,2292.000)(4062.000,2412.000)(4032.000,2292.000)
\path(4062,3012)(4062,5112)
\path(4092.000,4992.000)(4062.000,5112.000)(4032.000,4992.000)
\path(4062,5712)(4062,6012)
\path(4092.000,5892.000)(4062.000,6012.000)(4032.000,5892.000)
\path(4062,6612)(4062,6912)
\path(4092.000,6792.000)(4062.000,6912.000)(4032.000,6792.000)
\path(4062,7512)(4062,7812)
\path(4092.000,7692.000)(4062.000,7812.000)(4032.000,7692.000)
\path(4812,1212)(6387,1812)
\path(6285.541,1741.246)(6387.000,1812.000)(6264.182,1797.315)
\path(3312,1212)(1512,1812)
\path(1635.329,1802.513)(1512.000,1812.000)(1616.355,1745.592)
\path(762,2112)(762,2412)
\path(792.000,2292.000)(762.000,2412.000)(732.000,2292.000)
\path(762,3012)(762,3312)
\path(792.000,3192.000)(762.000,3312.000)(732.000,3192.000)
\path(762,3912)(762,4212)
\path(792.000,4092.000)(762.000,4212.000)(732.000,4092.000)
\path(762,4812)(3312,5412)
\path(3202.061,5355.313)(3312.000,5412.000)(3188.319,5413.718)
\path(7212,2112)(4812,5412)
\path(4906.843,5332.597)(4812.000,5412.000)(4858.319,5297.307)
\put(4062,312){\makebox(0,0)[b]{\smash{{{\SetFigFont{8}{14.4}{\rmdefault}{\mddefault}{\updefault}HOL}}}}}
\put(4062,1232){\makebox(0,0)[b]{\smash{{{\SetFigFont{8}{14.4}{\rmdefault}{\mddefault}{\updefault}Marvin\_}}}}}
\put(4062,1020){\makebox(0,0)[b]{\smash{{{\SetFigFont{8}{14.4}{\rmdefault}{\mddefault}{\updefault}lemmas}}}}}
\put(4062,2712){\makebox(0,0)[b]{\smash{{{\SetFigFont{8}{14.4}{\rmdefault}{\mddefault}{\updefault}Pointwise}}}}}
\put(7212,1812){\makebox(0,0)[b]{\smash{{{\SetFigFont{8}{14.4}{\rmdefault}{\mddefault}{\updefault}RefRels}}}}}
\put(762,1812){\makebox(0,0)[b]{\smash{{{\SetFigFont{8}{14.4}{\rmdefault}{\mddefault}{\updefault}Term}}}}}
\put(762,2712){\makebox(0,0)[b]{\smash{{{\SetFigFont{8}{14.4}{\rmdefault}{\mddefault}{\updefault}Pred}}}}}
\put(762,3612){\makebox(0,0)[b]{\smash{{{\SetFigFont{8}{14.4}{\rmdefault}{\mddefault}{\updefault}Command}}}}}
\put(762,4512){\makebox(0,0)[b]{\smash{{{\SetFigFont{8}{14.4}{\rmdefault}{\mddefault}{\updefault}State}}}}}
\put(4062,5412){\makebox(0,0)[b]{\smash{{{\SetFigFont{8}{14.4}{\rmdefault}{\mddefault}{\updefault}Execs}}}}}
\put(4062,6312){\makebox(0,0)[b]{\smash{{{\SetFigFont{8}{14.4}{\rmdefault}{\mddefault}{\updefault}RefCmd}}}}}
\put(4062,7212){\makebox(0,0)[b]{\smash{{{\SetFigFont{8}{14.4}{\rmdefault}{\mddefault}{\updefault}Recurse}}}}}
\put(4062,8112){\makebox(0,0)[b]{\smash{{{\SetFigFont{8}{14.4}{\rmdefault}{\mddefault}{\updefault}RefAssume}}}}}
\end{picture}
}
\end{center}
\caption{Structure of the RefLP theory
\label{fig:theorystruct}}
\end{figure}

Figure~\ref{fig:theorystruct} shows the structure of the theories
defined in \reflptool{}. 


{\tt Marvin\_lemmas}
extends the existing Isabelle/HOL theories
with additional lemmas required by the tool.
The theories {\tt Term} and {\tt Pred} model terms 
and predicates respectively, using a datatype to declare 
their syntax.
 The theory {\tt Command} models $Cmd$, representing the
commands in the wide-spectrum language; commands are defined using the
datatype facility.  
{\tt State} models program state as a set of bindings. 
{\tt Pointwise} models a
pointwise union and pointwise intersection operator, corresponding to
generalisations of $\lcup$ and $\lcap$. The theory {\tt RefRels}
models the polymorphic refinement relations ``refines to'' ($\refsto$)
and ``refinement equivalence'' ($\refeq$). {\tt Execs} models the type
$Exec$ and the function $\cexec$.  {\tt RefCmd} models
refinement of commands, from which a number of useful refinement laws
are derived.
{\tt Recurse} models the recursive features of the language. {\tt RefAssume}
extends the notion of command refinement to include assumption context.

The theories {\tt Command}, {\tt Execs} and {\tt RefCmd} are described
briefly below.

\subsection{Commands}

The theory {\tt Command} introduces the type {\tt Cmd},
corresponding to $Cmd$ defined in Section~\ref{reflp:wide-sp-l}.  Commands are
defined using Isabelle/HOL's datatype facility, and each command is
described via a type constructor and the type of its arguments.

\begin{isabelle}
\begin{tabular}{lcl}
datatype Cmd & == & $\lseq$ pred $\rseq$\\
           &  | &  $\{$ pred $\}$ \\
           &  | & Cmd $\pand$ Cmd \\
           &  | & Cmd $\por$ Cmd \\ 
           &  | & Cmd $\sand$ Cmd \\
           &  | & $\exists$ Var $@$ Cmd \\
           &  | & $\all$ Var $@$ Cmd \\
           &  | & fail \\
           &  | & abort \\
           &  | & skip \\
           &  | & PIdent(rterm)  
\end{tabular}
\end{isabelle}

\subsection{Executions}

The theory {\tt Execs} introduces the execution type $Exec$
and the function $\cexec$, mapping a command to an execution.
Firstly the type \isatext{StateMap} is defined, representing the set
of partial mappings between initial and final states.  

\begin{isabelle}
StateMap == State $\pfun$ State
\end{isabelle}
Partial
mappings are defined in Isabelle/HOL
in terms of total functions.
The labels \isatext{Some} and \isatext{None} are used to distinguish
defined and undefined values. More precisely, for a partial mapping
$f$ and element $x$, if $x$ is in the domain of $f$ then there exists
a $y$ such that \isatext{f(x) = Some y}, otherwise \isatext{f(x) = None}.
The label \isatext{the} can be used to strip away the label \isatext{Some}.

We define the type \isatext{Exec}, as a subtype of \isatext{StateMap},
using Isabelle's \isatext{typedef} facility. 
The set of executions corresponds to those partial functions satisfying the
three properties of executions
(\ref{exec-per-binding-dom}), (\ref{exec-constraining}) and
(\ref{exec-per-binding-range}) defined in Section~\ref{def-exec}:
\begin{isabelle}
  Exec == $\{$e: StateMap  $\cbar$ dom(e) = $\pset$ $\{$b. $\{$b$\}$ $\in$ dom(e)$\}$\\
\T2  $\land$ $\all$ s $\in$ dom(e). the e(s) $\subs$ s\\
\T2  $\land$ $\all$ s $\in$ dom(e). e(s) =  Some $\{$b. b $\in$ s \& (e($\{$b$\}$) = Some $\{$b$\}$)$\}\}$
\end{isabelle}

The function $\cexec$ is defined using primitive recursion. The
definition in Isabelle follows that given in
Figure~\ref{fig-cexec-basic}, except, for simplicity, definedness has
been dropped from the definition of $\cexec$ for specifications and
assumptions.  
We also include the procedure call semantics (defined in 
Section~\ref{def-proc-call}).
In the definition below,
\verb+PInt+ and \verb+PUn+
correspond to pointwise
intersection and pointwise union, and `\verb+o;+' corresponds to
composition of partial functions.
\begin{isabelle}
  "exec p $\lseq$ x $\rseq$ = ($\lambda$ s . Some (s $\inter$ bnds(x)))" \\
  "exec p $\{$ x $\}$ = ($\lambda$ s . if  s $\subs$ bnds(x)  then Some s else None)" \\
  "exec p (c1 $\pand$ c2) = ((exec p c1) $\lcap$ (exec p c2))" \\
  "exec p (c1 $\por$ c2) = ((exec p c1) $\lcup$ (exec p c2))" \\
  "exec p (c1 $\sand$ c2) = ((exec p c1) $\comp$ (exec p c2))" \\
  "exec p ($\pexists$ v $@$ c) = exists v (exec p c)" \\
  "exec p ($\pforall$ v $@$ c) = forall v (exec p c)" \\
  "exec p fail  = ($\lambda$ s. Some $\emptyset$)" \\
  "exec p abort = ($\lambda$ s. if s = $\emptyset$ then Some $\emptyset$ else None)"  \\
  "exec p skip  = ($\lambda$ s. Some s)" \\
  "exec p fid(t) = \\
\T1    (if fid: $\dom$(p)\\
\T1    then Rep\_Exec((the (p fid))(t))\\
\T1    else ($\lambda$ s. if s = $\emptyset$ then Some $\emptyset$ else None))"
\end{isabelle}

When processing a {\tt primrec} declaration, 
 Isabelle performs a number of checks.
 These checks include ensuring
that there is exactly one reduction rule for each constructor,
and that the variables on the right-hand side of each definition
are captured by any variables given on the left. In this way {\tt primrec}
declarations are a safe and conservative way of defining functions
on recursively defined data structures.

Each of the reduction rules in the {\tt primrec} declaration are added
to the default simplification set, and are
applied automatically when a simplification tactic is called. Thus, in
general, the user does not need to refer to the reduction rules
explicitly.

\subsection{Refinement rules}

The theory {\tt RefCmd} introduces the refinement relations
$\refsto$
and $\refeq$ for commands, defined in the context of an environment
$\rho$.
For example, the rule \lref{\existsmono} is represented in \reflptool{} as:
\begin{isabelle}
  env $\rho$ $\Vdash$ S $\refsto$ T  ==> env $\rho$ $\Vdash$ $\exists$ v S $\refsto$ $\exists$ v T
\end{isabelle}
Refinement laws proven within \reflptool{} include algebraic properties,
predicate lifting laws, mononotonicity laws, specification and
assumption laws, and laws which extend and use the assumption context.
In particular, the laws listed in Appendix A have all been proven
within \reflptool{}. The refinement laws have been used to refine
a number of programs in \reflptool{}.

\newcommand{\ignore}[1]{}
\ignore{
\subsection{Example}

The factorial example presented in Section~\ref{sec-fac-xmp} has been
developed using \reflptool{}. We begin by defining the theory
{\tt Factorial}. It includes an auxiliary function {\tt Fact}
with signature $\nat \tfun \nat$ using a recursive definition:
\begin{isabelle}
   Fact 0 = 1 \\
   Fact (Suc X) = (Fact (X * Suc X))
\end{isabelle}
We also define a number of functions in terms of the $\apply$ function, for
example we define the function {\tt fact} as follows:
\begin{isabelle}
   apply fact [X] = Fact X
\end{isabelle}
The refinement process begins by setting up the initial subgoal
$\Spec{V = fact(U)} \refsto {\tt ?X}$, 
where \verb+?X+ is an unknown variable to be instantiated to the
final program. This is represented in \reflptool{} as:
\begin{isabelle}
1. spec (varT V == funT fact [varT U]) [=(p) ?X
\end{isabelle}
 Note that currently \reflptool{} restricts the
set of values that a variable can take to the set of natural numbers.
Therefore the assumption $\Assert{u \in \nat}$ is redundant and
not included.
A step-wise refinement approach is used; to facilitate this
 we use the rule \lref{\refstotrans} to introduce intermediate programs.
\begin{isabelle}
by (resolve\_tac [refsto\_trans] 1); \\
1. spec (varT V == funT fact [varT U]) [=(p) ?T1 \\
2. ?T1 [=(p) ?X
\end{isabelle}
We now introduce the specification 
$\Spec{U = 0 \lor U > 0}$. Firstly we introduce a specification
containing the uninstantiated predicate $B$ using the
rule \lref{\introducespec}, however to apply this rule we must strengthen
the relation in the first subgoal using the law
\lref{\refeqstrongerrefsto}. These laws are combined using Isabelle's
{\tt RS} combinator, where {\tt thm1 RS thm2} resolves the conclusion of
{\tt thm1} with the first premise of {\tt thm2}.
\begin{isabelle}
by (resolve\_tac [introduce\_spec RS refeq\_stronger\_than\_refsto] 1); \\
1. |[?B]| \\
2. spec ?B pand \\
spec (varT V == funT fact [varT U]) [=(p) ?X
\end{isabelle}
Due to the representation of predicates as mappings from bindings to boolean
in \reflptool{} we cannot include just \verb+?B+ as a subgoal.
Instead we have the subgoal
\verb+|[?B]|+, which states that the predicate \verb+?B+ is true everywhere.
To complete the specification introduction we instantiate 
the unknown \verb+?B+ by matching against a lemma stating
$U = 0 \lor 0 < U$, represented in Isabelle as:
\begin{isabelle}
|[ varT U == funT zero [] || funT zero [] << varT U ]|
\end{isabelle}
We apply this lemma to replace \verb+?B+ by this new predicate:
\begin{isabelle}
by (res\_inst\_tac [("U","U")] factorial\_lemma1 1); \\
1. spec (varT U == funT zero [] || \\
~~~~  funT zero [] << varT U) pand \\
~~~~  spec (varT V == funT fact [varT U]) [=(p) ?X
\end{isabelle}
The refinement continues for a number of steps. Subcommands are refined
by explicitly calling the appropriate monotonicity laws. The refinement
follows the manual refinement up to the point of introducing the
recursive call; the resulting program is as follows:

\begin{isabelle}
(spec (varT U == funT zero []) sand spec (varT V == funT one [])) por \\
spec (funT zero [] << varT U) sand \\
exists UP \\ 
~~ (exists VP \\ 
~~~~   (spec (varT U == funT succ [varT UP]) sand \\
~~~~~~    spec (varT VP == funT fact [varT UP]) sand \\
~~~~~~    spec (varT V == funT mult [varT VP, varT U])))
\end{isabelle}
The number of steps required in \reflptool{} to get to this stage
is 75, compared with 11 in the manual example. The number of steps
could be reduced dramatically by implementing window inference
style tactics. Currently \reflptool{} does not support recursion,
parameters and procedures.

\subsection{Further work}

\reflptool{} is not yet complete. Further work includes:
\begin{itemize}
\item modelling parameterised commands, procedures and recursion blocks;
\item proof of the recursion introduction rule;
\item implementation of window inference tactics;
\item development of a user interface that uses a lighter syntax than
the direct encoding in Isabelle.
\end{itemize}
Upon completion of the first two, we will be able to complete the 
refinement of the factorial example. It is also our intention to
develop other examples in \reflptool{}.

}

\section{Conclusion}

We have presented a refinement calculus for logic programming.  The
calculus contains a wide-spectrum logic programming language,
including both specification and executable constructs.  We presented
a declarative semantics for this wide-spectrum language based on
executions, which are partial functions from states to states.  Part
of these semantics is a formal notion of refinement over programs in
the wide-spectrum language.
To illustrate the semantics we presented refinement laws proved in our
semantics, and included the refinement of a non-trivial program.  We
also discussed a tool that supports the refinement calculus, based on
the Isabelle theorem prover.

In earlier work \cite{HNS:REFLP}, we defined the meaning of a command $c$
in the wide-spectrum language by a pair of predicates:
$ok.c$ is a predicate defining the initial condition under which execution
of the command is well-defined (essentially representing the assumptions
the command makes about its context), while $ef.c$ is the effect of the
command, provided that its assumptions are satisfied.
The two semantics are closely related in that, for every command $c$,
\begin{zed}
    \pred{ok.c} = \stateuni \dom (\cexec c) \\
    \pred{ok.c \land ef.c} = (\cexec c)(\pred{ok.c})
\end{zed}
However, the earlier paper lacked a rigorous treatment of procedures,
parameters, and recursion, and the semantics we use here is chosen to
facilitate the presentation of those concepts.

In \cite{typesinv:ACSC00} we demonstrate how we represent types in
the refinement calculus via specifications and assumptions,
and include a wider discussion on contextual refinement.  
In \cite{datareft:fmp98} we introduce \emph{data refinement},
where we change the representation of the types of a procedure.  This
allows a specification type to be replaced by an implementation
type, or a procedure to be implemented by a more efficient representation.
Data refinement is
performed on a procedure-by-procedure basis.  We have three kinds of
data refinement, distinguished by interface issues.  
This work is extended
in \cite{modreft:LOPSTR00}, where we consider data refinement on groups of
procedures.  We introduce a notion of
\emph{module} into the wide spectrum language, and by restricting the set
of programs that may use a module, we can develop efficient
representations of the original types.
In \cite{codegen:02}
we present a prototype tool for generating Mercury \cite{Mercury:95}
code
from programs in our wide-spectrum language.
We define what it means for a wide-spectrum program to be executable,
and derive type and mode declarations by analysing the structure of the
program.

\ignore{
As current work we
are extending the semantics to include a notion of ``don't care''
non-determinism,
where a program may be
specified to return some, but not all, of its possible
solutions.
Such non-determinism may be approximated in Prolog via the use of green cuts.
We are also investigating
extending the semantics presented here by
introducing \emph{unifiers}
in our semantic domain, i.e., mappings from variables to terms.  
This will allow us to
discuss finitely representable states, and hence further bridge the gap
between our specification language and logic programming languages.
}

{\bf Acknowledgements.}
This work reported in this paper is supported by
Australian Research Council grant number A49937007: \emph{Refinement Calculus
for Logic Programming}.

\bibliographystyle{tlp}
\bibliography{biblio}

\appendix

\section{Refinement laws}
\label{sec:reflaws:listing}
\subsection{Algebraic laws}

\refrule{\pandcommute}{
  c_1 \pand c_2 \refeq c_2 \pand c_1
}
\refrule{\pandassoc}{
(c_1 \pand c_2) \pand c_3 \refeq c_1 \pand (c_2 \pand c_3)
}
\refrule{\pandidempotent}{c \pand c \refeq c}
\refrule{\porcommute}{c_1 \por c_2 \refeq c_2 \por c_1}
\refrule{\porassoc}{(c_1 \por c_2) \por c_3 \refeq c_1 \por (c_2 \por c_3)}
\refrule{\poridempotent}{
    c \por c \refeq c
}
\refrule{\sandassoc}{ (c_1 \sand c_2) \sand c_3 \refeq c_1 \sand (c_2 \sand c_3)}
\refrule{\sandidempotent}{ (c_1 \sand c_1) \refeq c_1 }
\refrule{\panddistrib}{
c_1 \pand (c_2 \por c_3) \refeq (c_1 \pand c_2) \por (c_1 \pand c_3)
}
\refrule{\pordistrib}{c_1 \por (c_2 \pand c_3) \refeq (c_1 \por c_2) \pand (c_1 \por c_3)}
\refrule{\leftsandoverpor}{c_1,(c_2 \por c_3) \refeq (c_1,c_2) \por (c_1,c_3)}
\refrule{\leftsandoverpand}{c_1,(c_2 \pand c_3) \refeq (c_1,c_2) \pand (c_1,c_3)}
\refrule{\pandoversand}{
   c_1 \pand (c_2 \sand c_3) \refsto (c_1 \pand c_2) \sand c_3}
\refrule{\rightsandoverpor}{
  (c_1 \por c_2) \sand c_3  \refeq  (c_1 \sand c_3) \por (c_2 \sand c_3) 
}
\refrule{\foralldistrib}{
(\forall X @ (c_1 \pand c_2)) \refeq (\forall X @ c_1) \pand (\forall X @ c_2)
}
\refrule{\existsdistrib}{
(\exists X @ (c_1 \por c_2)) \refeq (\exists X @ c_1) \por (\exists X @ c_2)}
\refrule{\pandtosand}{c_1 \pand c_2 \refsto c_1 \sand c_2 }
\refrule{\extendscopeexistsoverpand}{
\Rule{X \nfi c_2}{(\exists X @ c_1) \pand c_2 \refeq (\exists X @ c_1 \pand c_2)}}

\subsection{Refinement relation laws}

\refrule{\refstoreflex}{
c \refsto c}
\refrule{\refeqreflex}{
c \refeq c}
\refrule{\refstotrans}{
\Rule{c_1 \refsto c_2; c_2 \refsto c_3}{c_1 \refsto c_3}}
\refrule{\refeqtrans}{
\Rule{c_1 \refeq c_2; c_2 \refeq c_3}{c_1 \refeq c_3}}
\refrule{\refeqsymm}{
\Rule{c_1 \refeq c_2}{c_2 \refeq c_1}}
\refrule{\refstoantisymm}{
\Rule{c_1 \refsto c_2; c_2 \refsto c_1}{c_1 \refeq c_2}}
\refrule{\refeqstrongerrefsto}{
\Rule{c_1 \refeq c_2}{c_1 \refsto c_2}}

\subsection{Monotonicity laws}

\refrule{\pandmono}{
    \Rule{ c_1 \refsto c_2; c_3 \refsto c_4}
         { c_1 \pand c_3 \refsto c_2 \pand c_4}
}

\refrule{\pormono}{
    \Rule{ c_1 \refsto c_2; c_3 \refsto c_4}
         { c_1 \por c_3 \refsto c_2 \por c_4 }
}
\refrule{\sandmono}{
    \Rule{ c_1 \refsto c_2; c_3 \refsto c_4}
         { c_1 \sand c_3 \refsto c_2 \sand c_4 }
}

\refrule{\existsmono}{
    \Rule{c_1 \refsto c_2}
         {(\exists X @ c_1) \refsto (\exists X @ c_2)}
}

\refrule{\forallmono}{
    \Rule{c_1 \refsto c_2}
         {(\forall X @ c_1) \refsto (\forall X @ c_2)}
}

\subsection{Predicate lifting}

\refrule{\liftpand}{
  \LPSpec{P} \pand \LPSpec{Q} \refeq \LPSpec{P \land Q}
}

\refrule{\liftpor}{
  \LPSpec{P} \por \LPSpec{Q} \refeq \LPSpec{P \lor Q}
}

\refrule{\liftexists}{
  \exists X @ \LPSpec{P} \refeq \LPSpec{\exists X @ P}
}

\refrule{\liftforall}{
  \forall X @ \LPSpec{P} \refeq \LPSpec{\forall X @ P}
}

\subsection{Specification and assumption laws}

\refrule{\weakenassumpt}{
 \Rule{P \entails Q}{\Assert{P} \refsto \Assert{Q}}}
\refrule{\removeassumpt}{ \Assert{A} \sand c \refsto c }
\refrule{\combineassumpt}{\Assert{A} \sand \Assert{B} \refeq 
\Assert{A \land B}}
\refrule{\establishassumpt}{\LPSpec{P} \refeq \LPSpec{P} \sand \Assert{P}}
\refrule{\equivspec}{\Rule{P \equiv Q}{\Spec{P} \refeq \Spec{Q}}}
\refrule{\assumptafterspec}{
\Rule{P \entails A}{\Spec{P} \refeq \Spec{P} \sand \Assert{A}}}

\subsection{Context}

\refrule{\introduceassumpt}{\Rule{\Gamma \entails A}{\Gamma \entails (c \refeqpoint \Ass{A},c)}}
\refrule{\introducespec}{ \Rule{\Gamma \entails B}{\Gamma \entails (c \refeqpoint \Spec{B} \pand c)}}

\refrule{\assumptincontext}{
\Rule{\Gamma \land A \entails (c_1 \refstopoint c_2)}{\Gamma \entails (\Assert{A} \sand c_1 \refstopoint \Assert{A} \sand c_2)}
}
\refrule{\specincontext}{
 \Rule{\Gamma \land A \entails (c_1 \refstopoint c_2)}{\Gamma \entails (\Spec{A} \sand c_1 \refstopoint \Spec{A} \sand c_2)}
}
\refrule{\equivunderassumpt}{
\Rule{ A \entails (P \iff Q)}{\Assert{A} \sand \Spec{P} \refeq
\Assert{A} \sand \Spec{Q}}}
\refrule{\useparallelspec}{
    \Rule{I \entails (P \iff Q)}
    {\Spec{I} \pand \Spec{P} \refeq \Spec{I} \pand \Spec{Q}}
}
\refrule{\caseanalysis}{
\Rule{\Gamma \entails P \lor Q}{\Gamma \entails (c \refstopoint (\LPSpec{P} \sand c) \por (\Spec{Q} \sand c))}}

\end{document}